\documentclass[twocolumn]{pasj00}

\begin{document}
\SetRunningHead{Shidatsu et al.}{X-Ray and Near-Infrared Observations of GX 339--4}
\Received{2011/03/29}
\Accepted{2011/05/17}

\title{X-Ray and Near-Infrared Observations of GX 339--4 in the Low/Hard State with Suzaku and IRSF}

\author{Megumi \textsc{Shidatsu}\altaffilmark{1}, Yoshihiro \textsc{Ueda}\altaffilmark{1}, Fumie \textsc{Tazaki}\altaffilmark{1},
 Tatsuhito \textsc{Yoshikawa}\altaffilmark{1},\\ Takahiro \textsc{Nagayama}\altaffilmark{2},
Tetsuya \textsc{Nagata}\altaffilmark{1}, Nagisa \textsc{Oi}\altaffilmark{3}, Kazutaka \textsc{Yamaoka}\altaffilmark{4},\\
Hiromitsu \textsc{Takahashi}\altaffilmark{5},
Aya \textsc{Kubota}\altaffilmark{6}, Jean \textsc{Cottam}\altaffilmark{7}, Ronald \textsc{Remillard}\altaffilmark{8}, 
and Hitoshi \textsc{Negoro}\altaffilmark{9}}
\altaffiltext{1}{Department of Astronomy, Kyoto University, Kyoto 606-8502}
\email{shidatsu@kusastro.kyoto-u.ac.jp}
\email{ueda@kusastro.kyoto-u.ac.jp}
\altaffiltext{2}{Department of Astrophysics, Nagoya University, Aichi 464-8602}
\altaffiltext{3}{Department of Astronomy, Graduate University for Advanced Studies, Tokyo 181-8588}
\altaffiltext{4}{Department of Physics, Aoyama Gakuin University, Kanagawa 229-8558}
\altaffiltext{5}{Department of Physical Science, Hiroshima University, Hiroshima 739-8526}
\altaffiltext{6}{Department of Electronic Information Systems, Shibaura Institute of Technology, Saitama 337-8570}
\altaffiltext{7}{Exploration of the Universe Division, NASA Goddard Space Flight Center, Greenbelt, MD 20771, USA}
\altaffiltext{8}{Department of Physics, Massachusetts Institute of Technology, Cambridge, MA 02138, USA}
\altaffiltext{9}{Department of Physics, Nihon University, Tokyo 101-8308}

\KeyWords{accretion, accretion disks --- black hole physics --- 
infrared: stars --- stars: individual (GX 339--4)
 --- X-rays: binaries} 

\maketitle
\begin{abstract}

X-ray and near-infrared ($J$-$H$-$K_{\rm s}$) observations of the
Galactic black hole binary GX 339--4 in the low/hard state were
performed with Suzaku and IRSF in 2009 March. The spectrum
in the 0.5--300 keV band is dominated by thermal Comptonization of
multicolor disk photons, with a small contribution from a direct disk
component, indicating that the inner disk is almost fully covered by
hot corona with an electron temperature of $\approx$175 keV. The
Comptonizing corona has at least two optical depths,
$\tau \approx 1,0.4$. Analysis of the iron-K line profile yields an
inner disk radius of $(13.3^{+6.4}_{-6.0}) R_{\rm g}$ ($R_{\rm g} $
represents the gravitational radius $GM/c^2$), with the best-fit
inclination angle of $\approx50^\circ$. This radius is consistent with that
estimated from the continuum fit by assuming the conservation of
photon numbers in Comptonization.
Our results suggest that the standard disk of GX 339--4 is 
likely truncated before reaching the innermost stable circular orbit (for a
non rotating black hole) in the low/hard state at $\sim$1\% of the
Eddington luminosity.
The one-day averaged near-infrared light curves are found to be
correlated with hard X-ray flux with $F_{\rm Ks} \propto F_{\rm
X}^{0.45}$. 
The flatter near infrared $\nu F_{\nu}$ spectrum than the radio
one suggests that the optically thin synchrotron radiation from the
compact jets dominates the near-infrared flux.
Based on a simple analysis, we estimate the magnetic field and
size of the jet base to be $5\times10^4$ G and $6\times 10^8$ cm,
respectively. The synchrotron self Compton component is estimated to
be approximately 0.4\% of the total X-ray flux.

\end{abstract}

\section{Introduction}

Black hole binaries (BHBs), a class of the brightest Galactic X-ray
sources, are ideal objects to study the accretion disk physics around
a black hole. They exhibit distinct ``states'' with different
luminosity and spectral shape, which are mainly determined by the mass
accretion rate (see e.g., \cite{don07} for a recent review). When the
accretion rate is high (but below several 10\% of the Eddington
limit), the spectra of BHBs are dominated by a soft X-ray component
below 10 keV. This state is called the ``high/soft'' state, and its
dominant soft component is described by a Multi-Color Disk (MCD)
model, formed by a superposition of blackbody radiation at different
temperatures emitted from various radii of the disk \citep{mit84}. In
this state, many observations suggest that the disk is extending down
to the innermost stable circular orbit (ISCO).

In contrast, at lower accretion rates, the hard X-ray ($> 10$ keV)
contribution to the entire flux becomes more substantial. This state,
called the ``low/hard'' state, shows X-ray spectra approximated by a
power law of photon index $1.5-1.9$ with a high energy cut-off at
$\sim 100$ keV.  This emission is interpreted to be produced by
unsaturated Compton up-scattering by hot electrons of seed photons
emitted by the disk. The physical properties and geometry of the
corona in relation to the disk, however, remain to be
studied. While there is a consensus that an optically thick disk
is truncated before reaching the ISCO at low mass accretion rates
(e.g., \cite{esi97}), discussion at what Eddington fraction the disk
becomes truncated is still controversial (see below).

Radio to optical observations of BHBs have provided evidence for the
presence of stable jets in the low/hard state, while the jet is quenched in the
high/soft state \citep{fen99a}. By using radio interferometers, the
central compact core and jet-like knots are resolved in several BHBs in
the low/hard state \citep[and references therein]{fen01}. In the radio
region, the spectral energy distribution (SED) of BHBs exhibits an
almost flat profile, which is considered to be optically thick
synchrotron radiation from the jets, like those observed from active
galactic nuclei \citep{bla79}. This flat spectrum is thought to
extend to the sub-mm or infrared wavelengths, above which the
synchrotron emission becomes optically thin and eventually declines as a
power law with an index $>1$ (SED units). Thus, observations at the wavelengths
where the synchrotron luminosity from the jets peaks give key
information to understand the energetics of the jet.

GX 339--4 is a transient Galactic BHB, discovered in the early 1970s
\citep{mar73}. Although this object was observed for many years at
various wavelengths, its binary-system parameters are still uncertain
(see Section 2). The accretion disk geometry in the low/hard
state has been studied from X-ray spectra at different
luminosities. \citet{tom09} show that, at $\sim 0.001$ of the
Eddingtion luminosity ($L_{\rm Edd}$), the standard disk is truncated
at $> 35 R_{\rm g}$ ($R_{\rm g} \equiv GM/c^2$ is the gravitational
radius, where $G$, $M$, and $c$ are the gravitational constant, black
hole mass, and light velocity, respectively.) \cite{mil06} argue that
the disk extends to $\approx 4 R_{\rm g}$ in the bright low/hard state
at $\sim 0.06$ of the Eddington luminosity, based on the detection
of a broad iron-K emission line. Similar results are reported by
\citet{tom08} at 0.023$L_{\rm Edd}$ and 0.008$L_{\rm
Edd}$. \citet{don10}, however, re-analyze the same data used by
\citet{mil06}, and find that the spectra are strongly affected by
pile-up (multiple photon events in CCD). Extracting the simultaneous
data that are free from pile-up, they obtain a narrower iron-K line
profile that rather favors a truncated disk with an innermost radius
of $>6 R_{\rm G}$. Thus, the disk evolution of GX 339--4 in the
low/hard state as a function of luminosity is not firmly established
yet. GX 339--4 is also a persistent radio source in the low/hard
state, whose spectrum is flat or slightly increases according to the
frequency with the spectral index $\approx +$0.1--0.2
\citep{cor00}. In spite of the importance of the near-infrared data,
simultaneous multi-band observations in the $J$-$H$-$K$ bands and
X-rays have been very limited so far.

To establish the inner disk structure of BHBs in the low/hard state,
we observed GX 339--4 using the X-ray satellite Suzaku in 2009
March, when it became active and stayed in the low/hard
state. Quasi-simultaneous near-infrared observations were performed
with the Infrared Survey Facility (IRSF) 1.4 m telescope
to reveal the jet activity. In this paper, we first summarize the
constraints on the system parameters of GX 339--4 based on previous
observations in Section 2. The Suzaku observations and data
reduction are described in Section 3, and analysis of the X-ray data
is given in Section 4. In Section 5, we present details of the IRSF
observations and data analysis. We discuss our X-ray and near-IR
results by comparing with previous observations of GX 339--4 and other
BHBs in Section 6. The conclusions are summarized in Section 7.

\section{Summary of Constraints on Inclination and Distance of GX 339--4}

\begin{figure*}[tbp]
\begin{minipage}{0.5\hsize}
\begin{center}
\FigureFile(70mm,){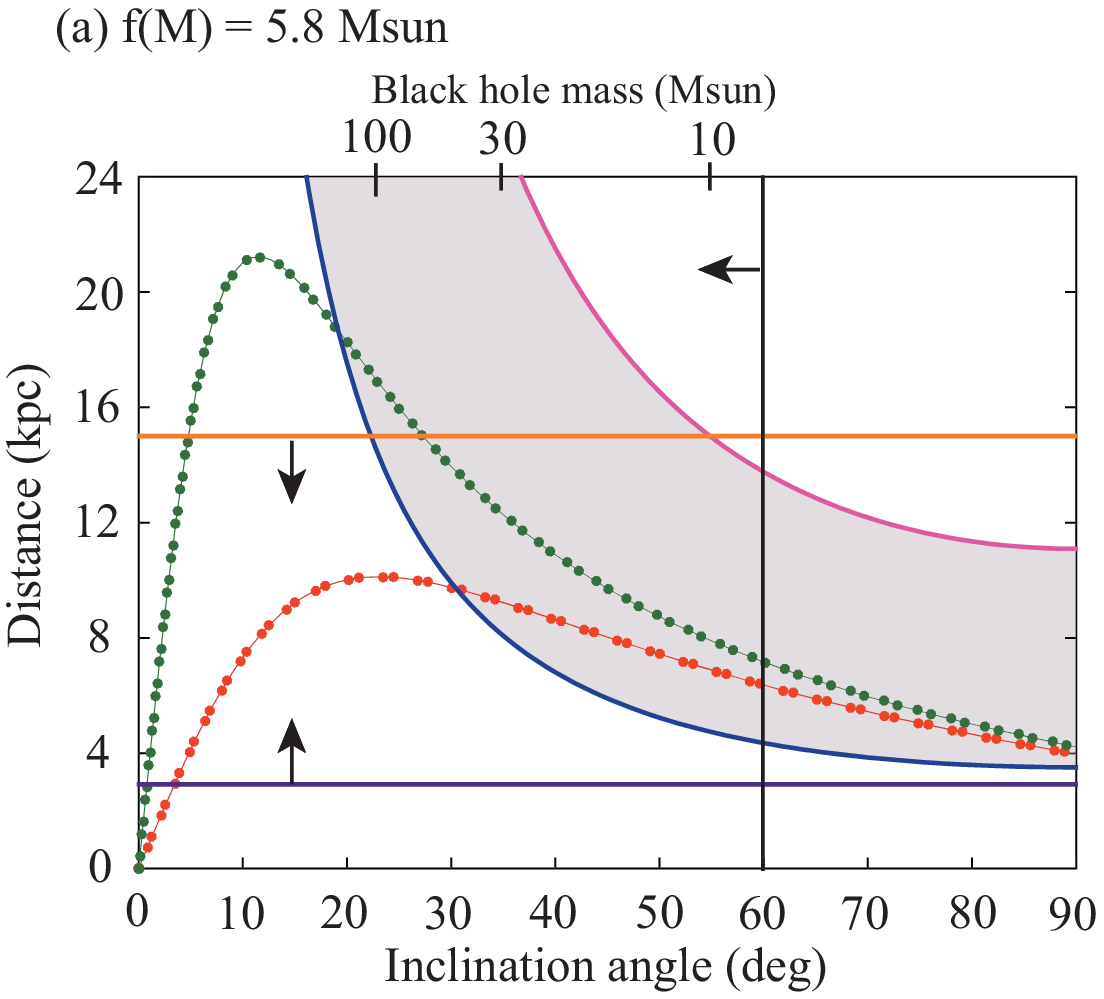}
\end{center}
\end{minipage}
\begin{minipage}{0.5\hsize}
\begin{center}
\FigureFile(70mm,){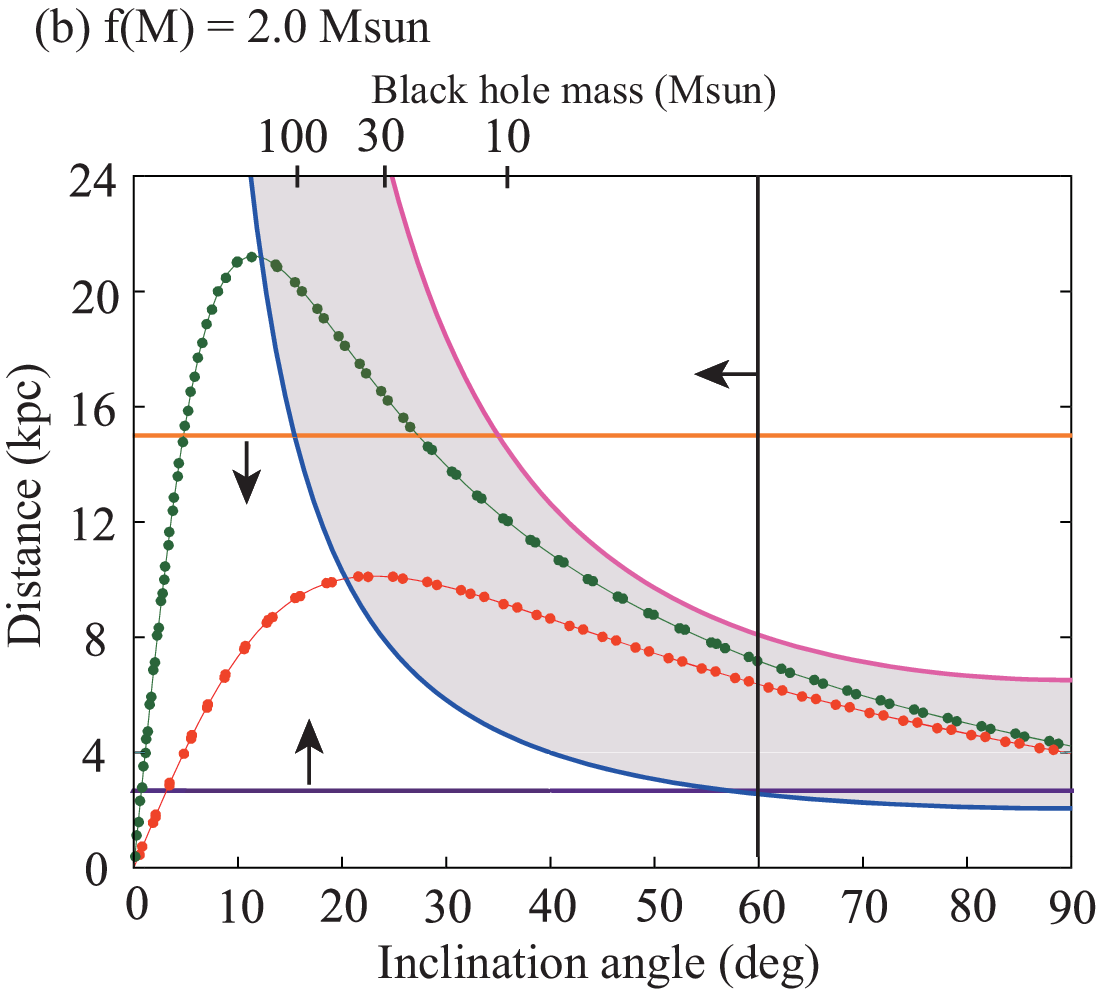}
\end{center}
\end{minipage}
\caption{Relation of the inclination angle, black hole mass, and
distance of GX 339--4. The abscissa represents the inclination angle
(bottom) and black hole mass (top).  The mass function is assumed to
be 5.8 $M_{\solar}$ in the left panel and 2.0 $M_{\solar}$ in the right
panel. The orange horizontal line shows the upper limit of distance (15 kpc)
derived from the velocity dispersions \citep{hyn04}, and the purple
line corresponds the lower limit (3 kpc) determined with the
extinction study \citep{zdz98}. The vertical line at $i=60^\circ$
shows the upper limit of the orbital inclination angle by the absence
of eclipse \citep{cow02}. The relation between the inclination angle
and distance, calculated from the 0.5--200 keV flux of GX 339--4 in
very high state \citep{yam09} is presented by assuming $L/L_{\rm
Edd}=1$ (pink, upper curve) and $0.1$ (dark blue, lower curve), with
the area for $0.1\leq L/L_{\rm Edd} \leq 1$ filled in gray. The red
and green dotted lines show the relation between the inclination
and distance derived from the jet motion observed by \citet{gal04},
where the intrinsic velocity is assumed to be 0.92$c$ and 0.98$c$,
respectively. \label{incl_D} }
\end{figure*}

The system parameters of GX 339--4 including the distance, inclination
angle, black hole mass (hereafter $M_{\rm BH}$), and the mass of
companion star ($M_{\rm C}$) are still unknown. Figure~\ref{incl_D}
summarizes the constraints derived from previous observations in the
inclination (or black hole mass) versus distance plane. The limit that
$i$ is less than $\sim 60^\circ$ is imposed by the fact that GX 339--4
exhibits no eclipse \citep{cow02}. The upper and lower limits of the
distance $D$ are estimated in several different ways. \citet{zdz98} found
that $D>3$ kpc by investigating the interstellar reddening, while
\citet{hyn04} estimated 6 kpc $<D<$ 15 kpc using the structure of
Na\emissiontype{D} line obtained through high-resolution
spectroscopy. \citet{zdz04} put the limit 7 kpc $<D<$ 9 kpc, using the
apparent radius of the secondary low-mass star based on optical and
infrared observations. In~Figure \ref{incl_D}, we plot the \citet{hyn04}
and \citet{zdz98} results for the upper and lower limits of the
distance, respectively.

The dependence of $M_{\rm BH}$ on the inclination angle is 
obtained from the mass function, which is expressed as
\begin{equation}
f(M_{\rm BH})=\frac{M_{\rm BH}}{(1+\frac{M_{\rm C}}{M_{\rm BH}})^2}
\sin^3 i = \frac{P K_{\rm C}^3}{2 \pi G}, \label{eq0}\end{equation}
where $P$, $G$, and $K_{\rm C}$ are the orbital period, the
gravitational constant, and the radial velocity semi-amplitude of the
companion star, respectively. \citet{hyn03} estimated $f(M_{\rm BH})$
of GX 339--4 to be 5.8 $M_{\solar}$ (2.0 $M_{\solar}$ at a 95\% confidence
lower limit) by observing the Doppler shifts of the N\emissiontype{III} and
He\emissiontype{II} emission lines from the irradiated companion star.
Ignoring the $M_{\rm C}/M_{\rm BH}$ term in Equation~\ref{eq0}, whose
value is less than 0.08 according to \citet{zdz04}, we calculate the
black hole mass corresponding to each inclination angle, which is
indicated in the upper axis. 

For a given $f(M_{\rm BH})$ value, we can also constrain the relation between
the distance and inclination angle, assuming that the fraction of the
Eddington luminosity in the very high state should be in the range of
$0.1\lesssim L/L_{\rm Edd} \lesssim 1$. Here we refer to the 0.2--100
keV luminosity in the very high state obtained with Suzaku in
2007 \citep{yam09}, $3.8\times 10^{38} (D/8 \mbox{ kpc})^2$ erg
s$^{-1}$.

Using the proper motion of the relativistic jets emitted from the
system, we are also able to connect the distance and inclination angle
(assumed to be perpendicular to the jet axis) for an assumed intrinsic
speed of the jet. In 2002, bright radio knots are observed from GX
339--4 at the state transition and the head of one knot moved 12
arcsec in approximately 300 days \citep{gal04}. From this apparent
velocity, we derive the relation between the inclination angle and
distance, assuming the intrinsic speed of the jet to be 0.98$c$
\citep{fen99b} or 0.92$c$ \citep{mir94}, a typical value observed from
the microquasar GRS 1915+105.

Also, recent work sets the lower limit of the companion-star mass,
$M_{\rm C} \geq 0.166 M_{\solar}$, by considering the mass transfer
process of low mass X-ray binaries \citep{mun08}. This minimum mass
makes a tight constraint on the binary inclination, $i>45$, by
assuming that the black hole mass is less than $15 M_{\solar}$, 
which is the largest mass ever known among Galactic BHBs (discussed in
\cite{kol10}). This is quite consistent with the constraints
summarized in Figure~\ref{incl_D} for both cases of the two $f(M_{\rm
BH})$ values.

\section{X-Ray Observations and Data Reduction}

\subsection{Observations}

\begin{figure}
\begin{center}
\FigureFile(70mm,){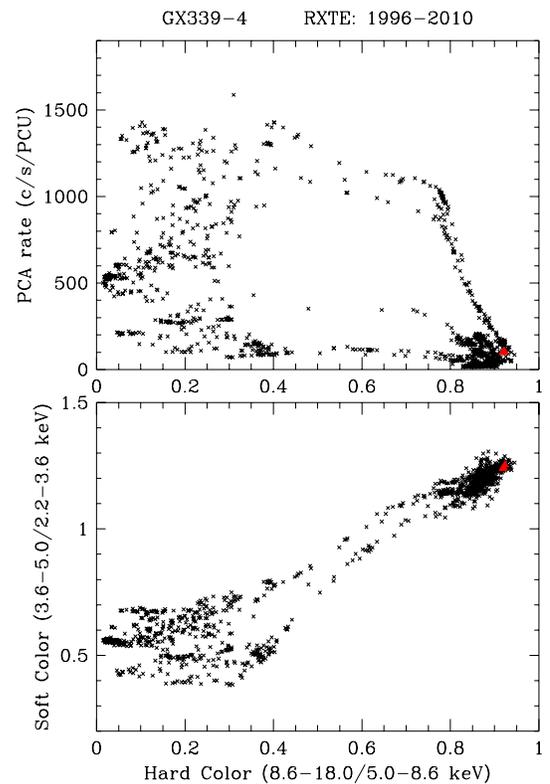}
\end{center}
\caption{
Hardness-intensity diagram (upper) and color-color diagram (lower) of
GX~339--4, obtained with RXTE/PCA during 1996--2010. The four
orbits during our Suzaku campaign is highlighted in red (filled triangle), one
point on MJD 54914 and three on MJD 54916.
\label{asm_color}}
\end{figure}

\begin{table*}
\caption{Summary of the Suzaku observations\label{tbl-1}}
\begin{center}
\begin{tabular}{ccccccc}
\hline
Label &  \multicolumn{2}{c}{Observation Date (UT)} & Observation ID & 
\multicolumn{3}{c}{Net Exposure (ksec)} \\ \cline{2-3} \cline{5-7}
 & Start & End & & XIS-0 and -3\footnotemark[$*$] & XIS-1\footnotemark[$\dagger$] & HXD \\ \hline

Epoch-1 & 2009 March 18 01:53:13 & 2009 March 18 23:04:19 & 403011010 & 21.4 & 3.9 & 35.4 \\
Epoch-2 & 2009 March 25 08:14:00 & 2009 March 26 06:30:24 & 403011020 & 19.4 & 3.6 & 34.8 \\
Epoch-3 & 2009 March 30 11:27:27 & 2009 March 31 12:31:10 & 403011030 & 19.6 & 3.6 & 35.1 \\
\hline
\multicolumn{7}{@{}l@{}}{\hbox to 0pt{\parbox{180mm}{\footnotesize
\footnotemark[$*$] XIS-0 and -3 were operated in the 1/4 window mode with 0.5 sec burst option.
\par\noindent
\footnotemark[$\dagger$] XIS-1 was in the full window mode with 0.5 sec burst option.
}\hss}}
\end{tabular}
\end{center}
\end{table*}

We performed three sequential ToO (Target of Opportunity) observations of
GX 339--4 with Suzaku \citep{mit07} on 2009 March 18, 25--26,
and 30--31 (hereafter Epoch-1, -2, and -3, respectively), each for a
net exposure of $\sim$40 ksec. The ToO observations were triggered
after the increase of the X-ray flux detected by RXTE/PCA
scanning observations in 2009 February \citep{mar09} and by Swift/BAT 
monitoring in March \citep{sti09}, based on our approved
Suzaku AO-3 program. 

In Figure~\ref{asm_color} we show the Suzaku hard-state observations in the
context of the large data archive for GX339--4 produced with the RXTE PCA
Instrument. We extracted the normalized PCA count rates in four
energy bands, as described by \citet{lin07}, to construct the
hardness-intensity diagram (HID; top panel) and color-color diagram
(bottom panel). Each of the 1159 plotted points represents an interval
of continuous exposure, with an average exposure time of 2.0 ks.
These data have been filtered to eliminate exposures obtained when the
source flux is fainter than 12.5 count sec$^{-1}$ PCU$^{-1}$ (or 5 mCrab at 2--30 keV),
so that the statistical error bars remain similar to or smaller than
the size of the plot symbols.

Four of these RXTE exposures overlap with the Suzaku observations (one
on MJD 54914 and three on 54916), and the results for those data are
plotted as red triangles. The HID in Figure \ref{asm_color} extends the results
shown by \citet[Figure 9d]{rem06} to cover RXTE
observations through 2010 July 8. In the earlier work, the HID points
are symbol-coded to show state assignments based on quantitative
definitions, and observations in the hard state completely surround
the regions occupied by the red points (i.e., the Suzaku-associated
exposures) shown in Figure \ref{asm_color}.

The Suzaku observations of GX 339--4 were conducted by using the
X-ray Imaging Spectrometers (XISs) and the Hard X-ray Detector (HXD)
at the HXD nominal position. The XISs are composed of three
frontside-illuminated (FI) CCDs (XIS-0, -2 and -3) and a
backside-illuminated (BI) CCD (XIS-1) covering the energy range of
0.2--12 keV. The BI-CCD has larger effective area than FI-CCDs below
$\approx$1.5 keV and is more sensitive to low energy photons. The HXD
consists of silicon PIN diodes and GSO crystal scintillators, and is
sensitive to hard X-ray photons in 10--70 keV and 40--600 keV bands,
respectively.  In our observations, the editing mode of the FI-CCDs
was $2 \times 2$ and that of the BI-CCD was $3 \times 3$.  To reduce
photon pile-up, XIS-0 and XIS-3 were operated with the 1/4 window and
0.5 sec burst option, and XIS-1 was used with the full window and 0.5
sec burst option. XIS-2 was not usable then because of the trouble of
the instrument on 2006 November 9. The summary of the observations is
given in Table~\ref{tbl-1}.

\subsection{Data Reduction}

\begin{figure*}
\begin{center}
\FigureFile(140mm,){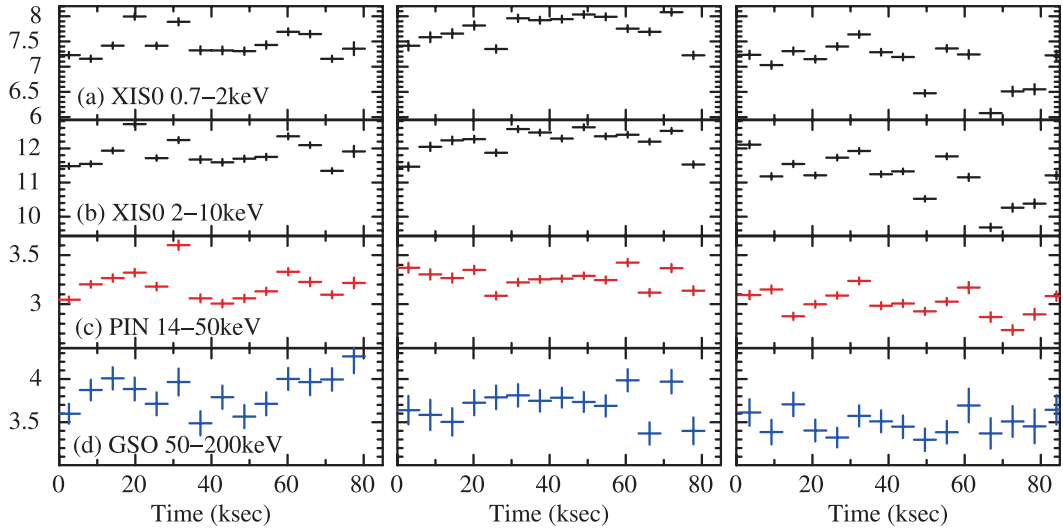}
\end{center}
\caption{X-ray light curves of GX 339--4 binned in 5760 sec. The
ordinates represent the count rate (counts sec$^{-1}$) in the energy range of
(a) XIS-0 0.7--2 keV, (b) XIS-0 2--10 keV, (c) PIN 14--50 keV, and (d)
GSO 50--200 keV. The left, middle, and right panels present the
results of Epoch-1, -2, and -3, respectively. 
The background subtraction (for the XIS and HXD) and dead time
correction (for the HXD) are applied.
\label{lc_suzaku}}
\end{figure*}

We utilized data products of both instruments reduced through Suzaku
pipeline processing version 2.3.12.25. To extract light curves
and spectra, the data were analyzed with HEAsoft version 6.6.3 and the
calibration data files (CALDB) released on 2009 October 7.  Because
the Suzaku attitude was still unstable shortly after the
maneuver to the target, the data acquired in the first 2 ksec in
Epoch-1 were discarded to avoid difficulties of analysis. We analyzed
the XIS cleaned event files in the ``sky coordinates'' system,
where the small attitude variation of the spacecraft during the orbit
is corrected. We extracted the counts of the FI-CCDs (XIS-0 and -3)
and the BI-CCD (XIS-1) from circular regions centered on the target
with the radii of $1.9'$ and $3.7'$, respectively. The regions
enclose $\sim$90\% of the available photons in each field of view,
whose area is a factor of 4 smaller in FI-CCDs than in BI-CCD because
of the window option. The background of the XIS were taken from a
source-free area with the same radii as the target. For the HXD, we
used the modeled background files produced by the Suzaku HXD
team\footnote{\tt
http://www.astro.isas.jaxa.jp/suzaku/analysis/hxd/pinnxb, gsonxb}.
The PIN background, composed of the non X-ray background (NXB) and the
cosmic X-ray background (CXB), was subtracted from the source spectra.
Although there is additional background contribution from the Galactic
ridge emission, the effect is estimated to be negligible ($\approx
0.5\%$ of the target flux) in our PIN data. For the background of GSO,
only the NXB component was considered because the contribution of the
CXB to the total background rate is less than 0.1\%
\footnote{http://www.astro.isas.ac.jp/suzaku/analysis/hxd/gsonxb/}.

The light curves of GX 339--4 in four energy bands, 0.7--2 keV, 2--10
keV, 14--50 keV, and 50--200 keV are plotted in Figure~\ref{lc_suzaku} for
the three epochs. The PIN and GSO light curves were corrected for dead
time, and the NXB was subtracted. Background subtraction was also
performed for the XIS data. As presented in Figure~\ref{lc_suzaku}, the
count rates of the target in these energy bands exhibit a similar
behavior to one another, suggesting that the overall spectral shape is
mostly constant.

\begin{figure}
\begin{center}
\FigureFile(70mm,){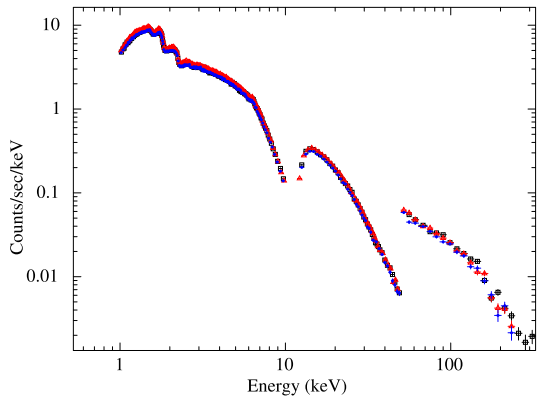}
\end{center}
\caption{
Time-averaged Suzaku spectra obtained in Epoch-1
(black, open square), -2 (red, open triangle) and -3 (blue, filled circle). 
For clarity the XIS-1 data are not plotted.\label{spec123}}
\end{figure}

Figure \ref{spec123} shows the time-averaged spectra of the XIS and the
HXD in the three observations. To improve statistics, XIS-0 and -3
were combined each other. We created the response matrix files and
auxiliary response files of the XIS with the FTOOLS {\tt xisrmfgen}
and {\tt xissimarfgen}, respectively. As the response files of the
HXD, we used {\tt ae\_hxd\_pinhxnome5\_20080716.rsp} for PIN, and {\tt
ae\_hxd\_gsohxnom\_20080129.rsp} and {\tt ae\_hxd\_\\gsohxnom\_crab\_20070502.arf}
\footnote{http://www.astro.isas.jaxa.jp/suzaku/analysis/hxd/gsoarf/} for GSO.
To include possible uncertainties associated with the calibration of the instruments,
1\% systematic error is introduced for each spectral bin of the XIS and the HXD.

\begin{figure}
\begin{center}
\FigureFile(70mm,){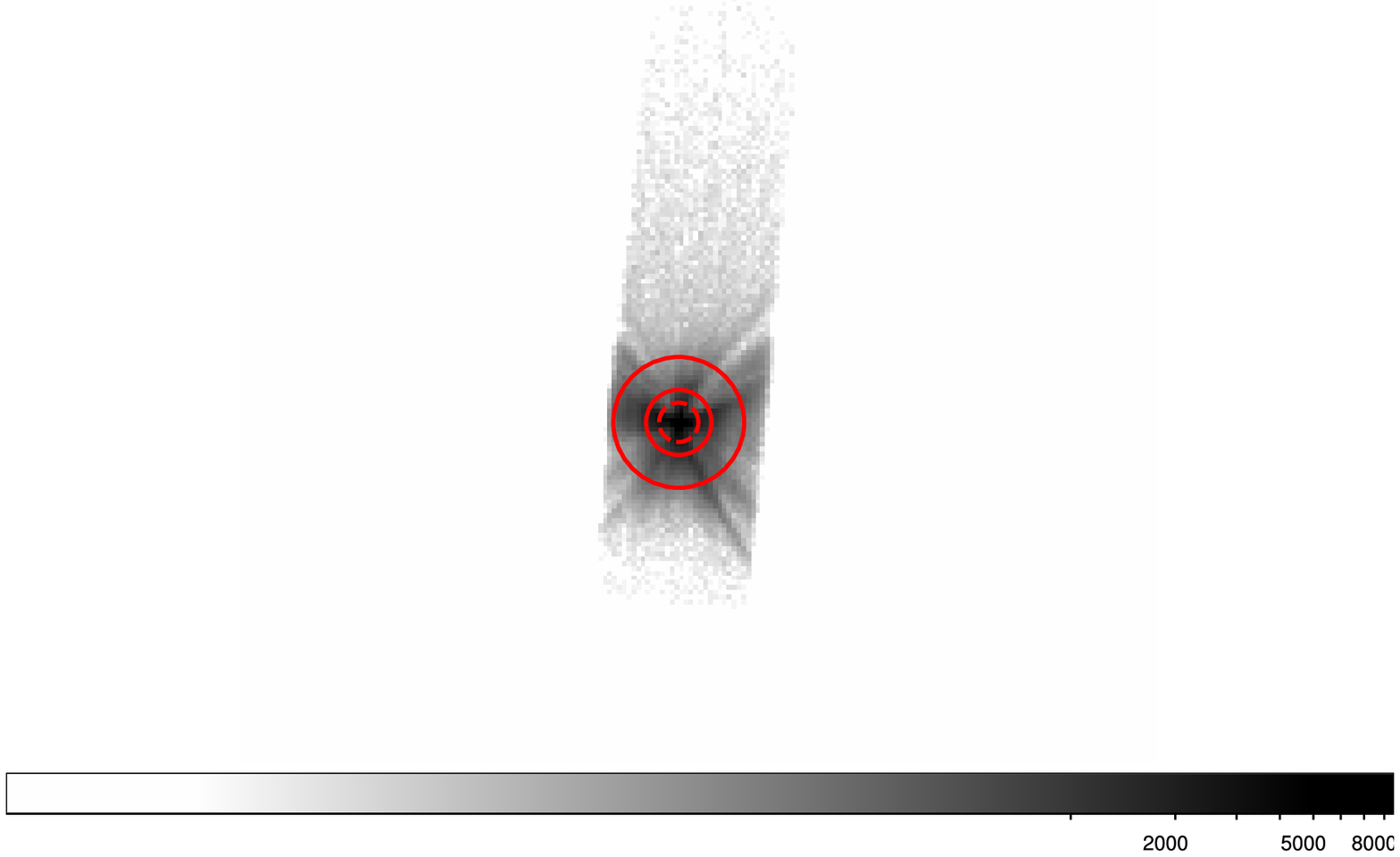}
\end{center}
\caption{
XIS-0 image of Epoch-1 in the sky coordinates.
The circles (red) correspond to radii of $1.9'$, $0.95'$, and $0.57'$ (dashed), which 
define the annulus regions used to check the pile-up effects.
\label{reg_img}}
\end{figure}

\begin{figure}
\begin{center}
\FigureFile(70mm,){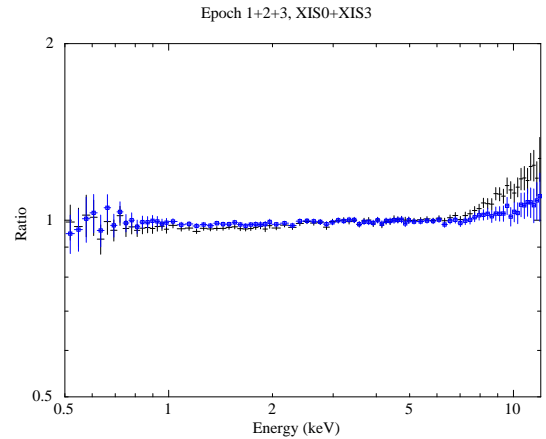}
\end{center}
\caption{
Ratios of the XIS-0$+$XIS-3 spectra (sum of Epochs 1, 2, and 3)
extracted from the whole circular region ($r<1.9'$, black) and annulus
of $r=0.57'-1.9'$ (blue, open square), divided by that from the
$r=0.95'-1.9'$ annulus. They are normalized to be unity at 5.0 keV.
\label{pileup}}
\end{figure}

To examine possible effects of pile-up in the XIS data, which
should be more significant in the core region of the point spread
function of the X-ray telescope, we also make the spectra of XIS-0 and
XIS-3 in two annuli whose inner radii are set to be 30\% ($0.57'$) and
50\% ($0.95'$) of the outer radius ($1.9'$). The regions are
illustrated in Figure~\ref{reg_img} for XIS-0. For each annulus
region, both XIS-0 and XIS-3 spectra in the three epochs are combined
into a single one. Figure~\ref{pileup} plots the ratios of the spectra
extracted from the whole circular region ($r<1.9'$, black) and the
$r=0.57'-1.9'$ annulus (blue), divided by that from the $r=0.95'-1.9'$
annulus, which can be regarded as free from pile-up. As noticed from
the figure, the spectral ratio for the whole circular region shows a
significant excess above $\sim$8 keV attributable to pile-up. Thus,
while we adopt the $r<1.9'$ spectra to achieve the best statistics,
the energy range above 8 keV is not utilized in our spectral
analysis. The difference from the $r=0.95'-1.9'$ spectrum is very small below
this energy band, and we confirm that our results on the innermost
disk radius derived from the iron-K profile (Section~4.3) are not
affected by pile-up. We find that its estimate from the continuum fit
(Section~4.2) could be affected by $\sim$20\% in Epoch~1 and by $<
10\%$ in Epochs~2 and 3, but this uncertainty does not change our
conclusion on the black hole mass presented in Section~4.3.

\section{Analysis \hspace{-0.2mm}of \hspace{-0.2mm}the \hspace{-0.2mm}Time-Averaged \hspace{-0.2mm}Suzaku \hspace{-0.2mm}Spectra}

In this section, we present the analysis of the Suzaku spectra
extracted above. The spectral fit is performed by using XSPEC version
11.3.2ag. In all spectral models, we assume solar abundances given by
\citet{and89}. We employ the {\tt wabs} model for Galactic
interstellar absorption, whose column density is set to be a free
parameter. The quoted errors refer to 90\% confidence levels for a
single parameter.

\subsection{Broad Band Fitting}

\begin{figure*}
\begin{center}
\FigureFile(55mm,){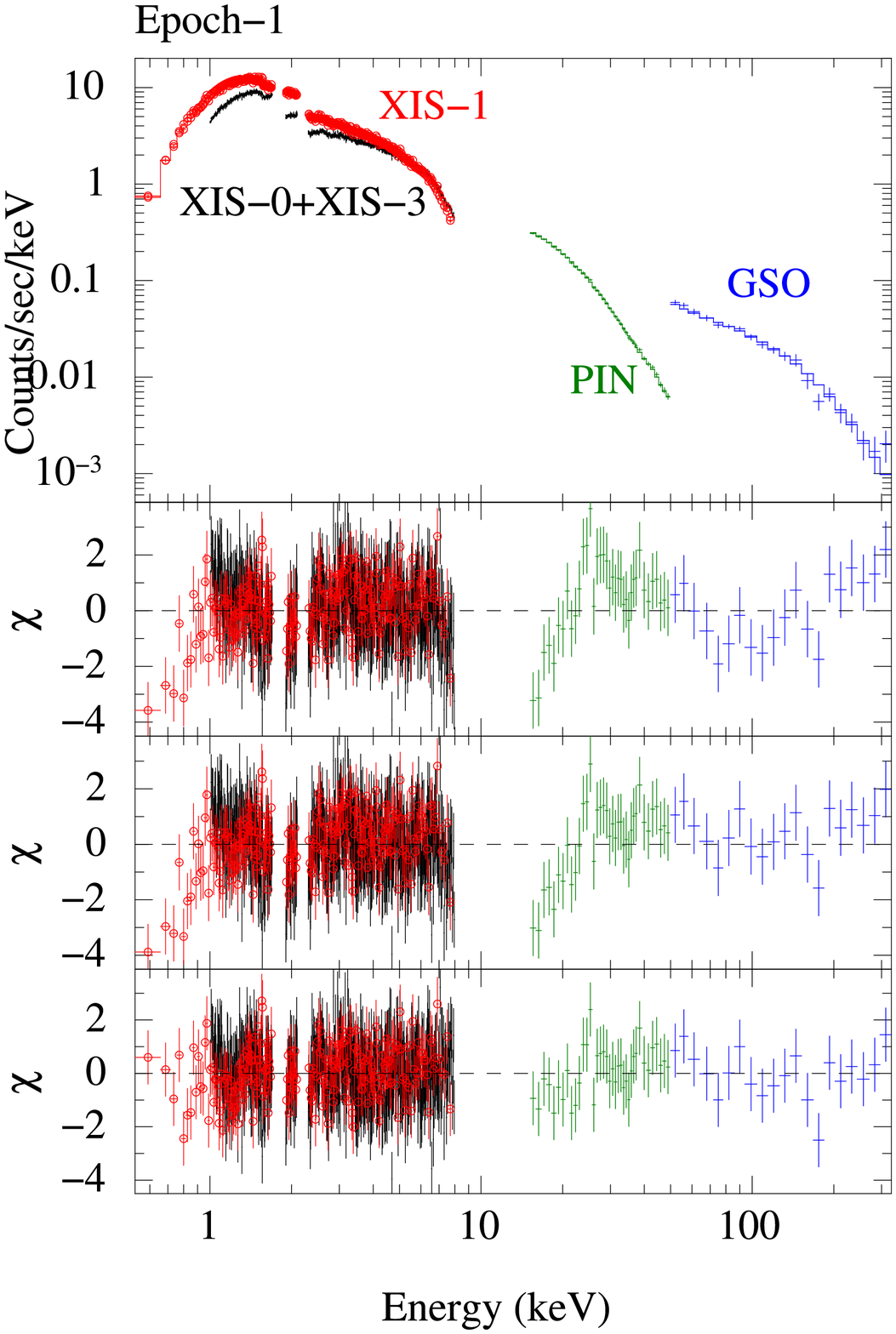}
\FigureFile(55mm,){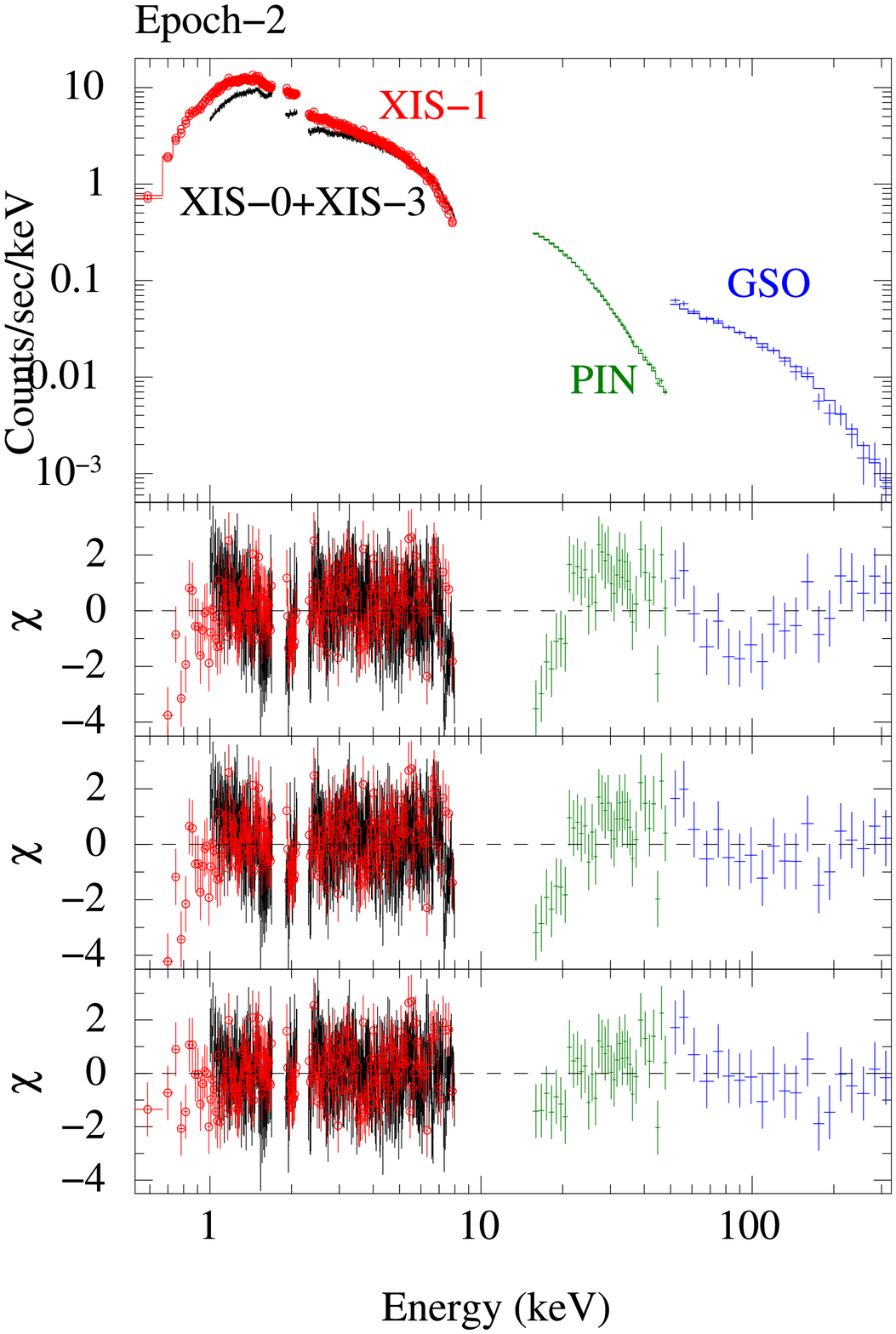}
\FigureFile(55mm,){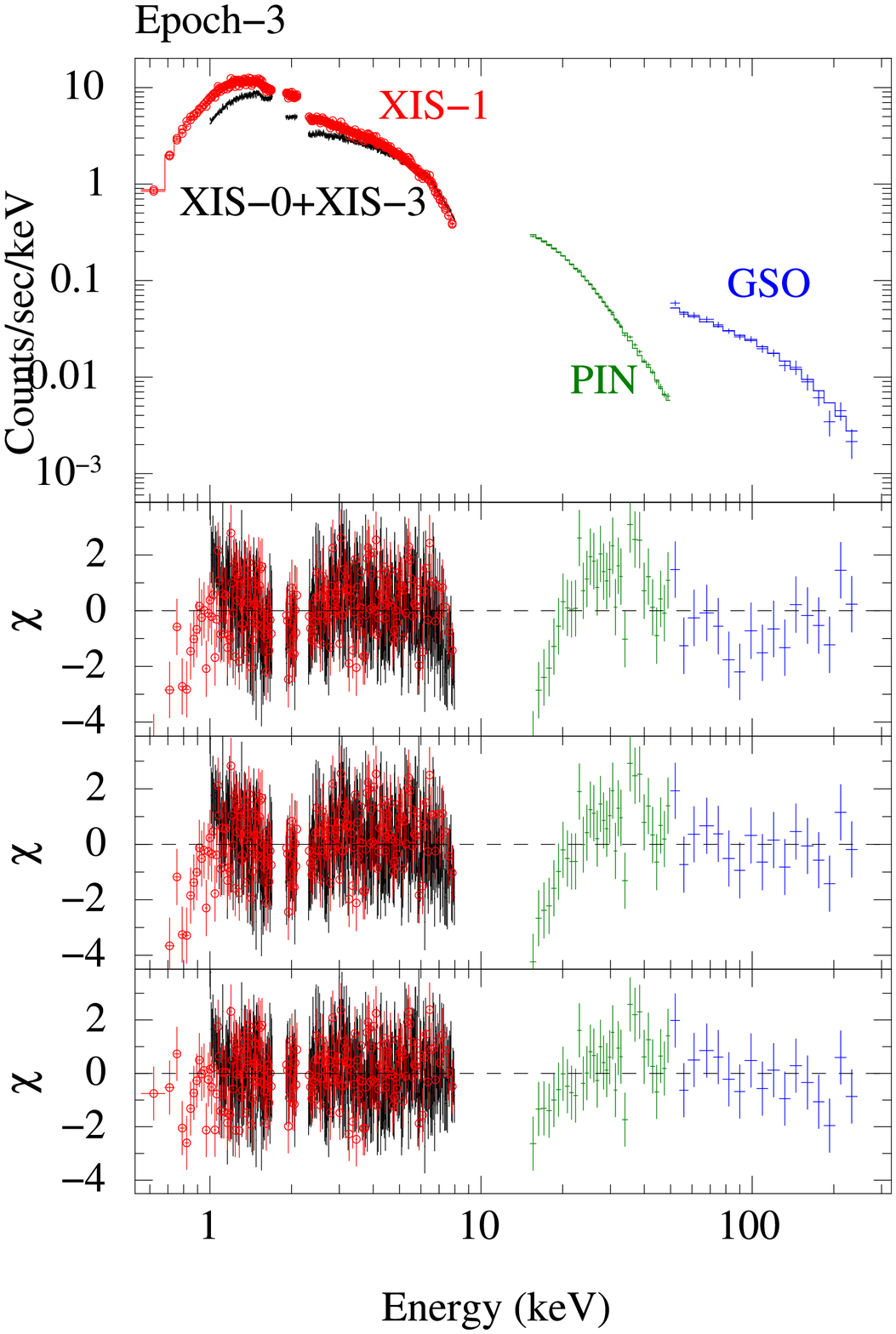}
\end{center}
\caption{Simultaneous fit to the XIS-1 (red, open circle),
XIS-0$+$XIS-3 (black), PIN (green), and
GSO (blue) spectra with the double {\tt compPS} model in Epoch-1, -2, and -3. 
The second, third, and bottom panels show the residuals of the 
fit with the single {\tt compPS} model without reflection, that 
with reflection, and the double {\tt compPS} model with reflection, 
respectively.\label{fit_comp}}
\end{figure*}

\begin{figure*}
\begin{center}
\FigureFile(70mm,){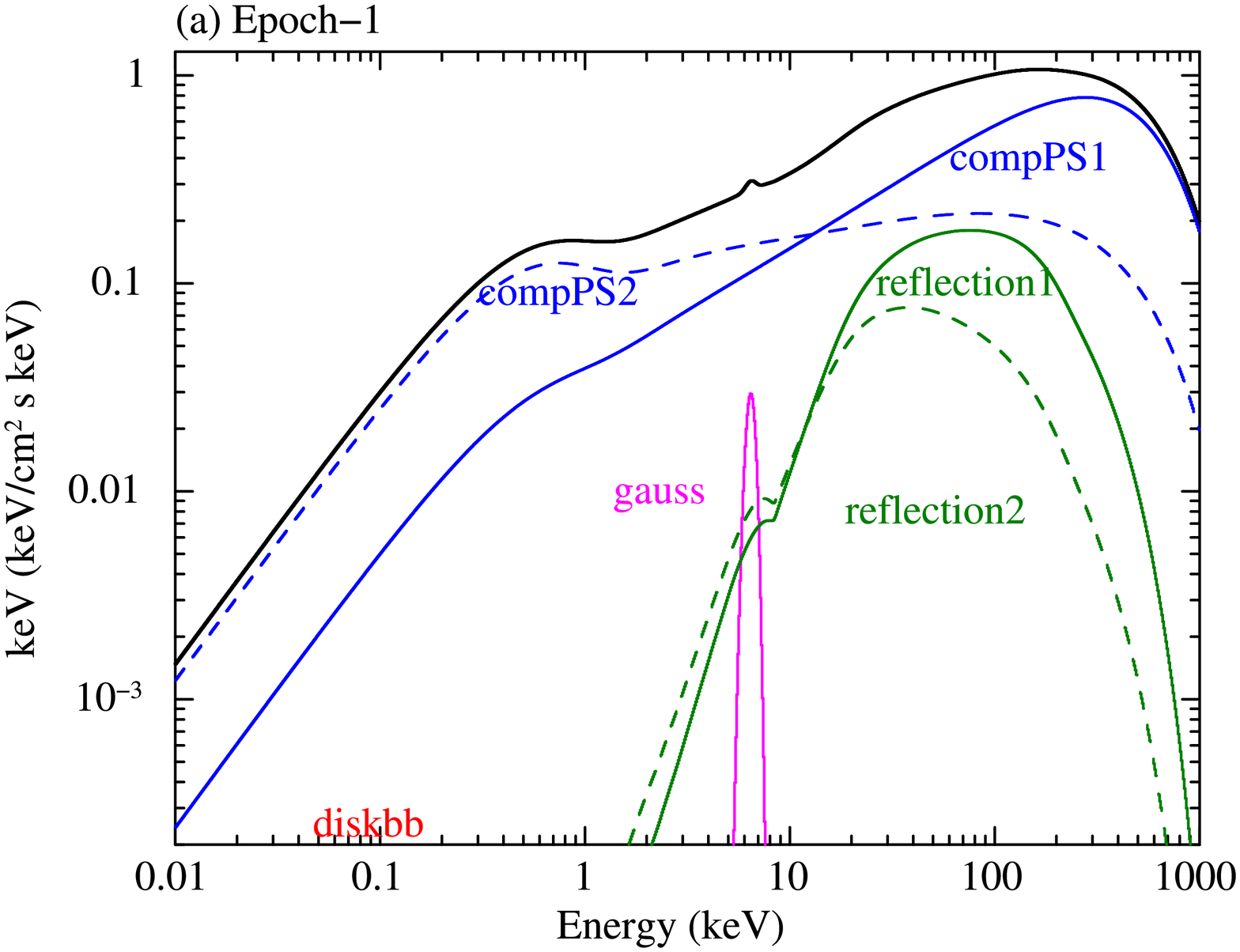}
\FigureFile(70mm,){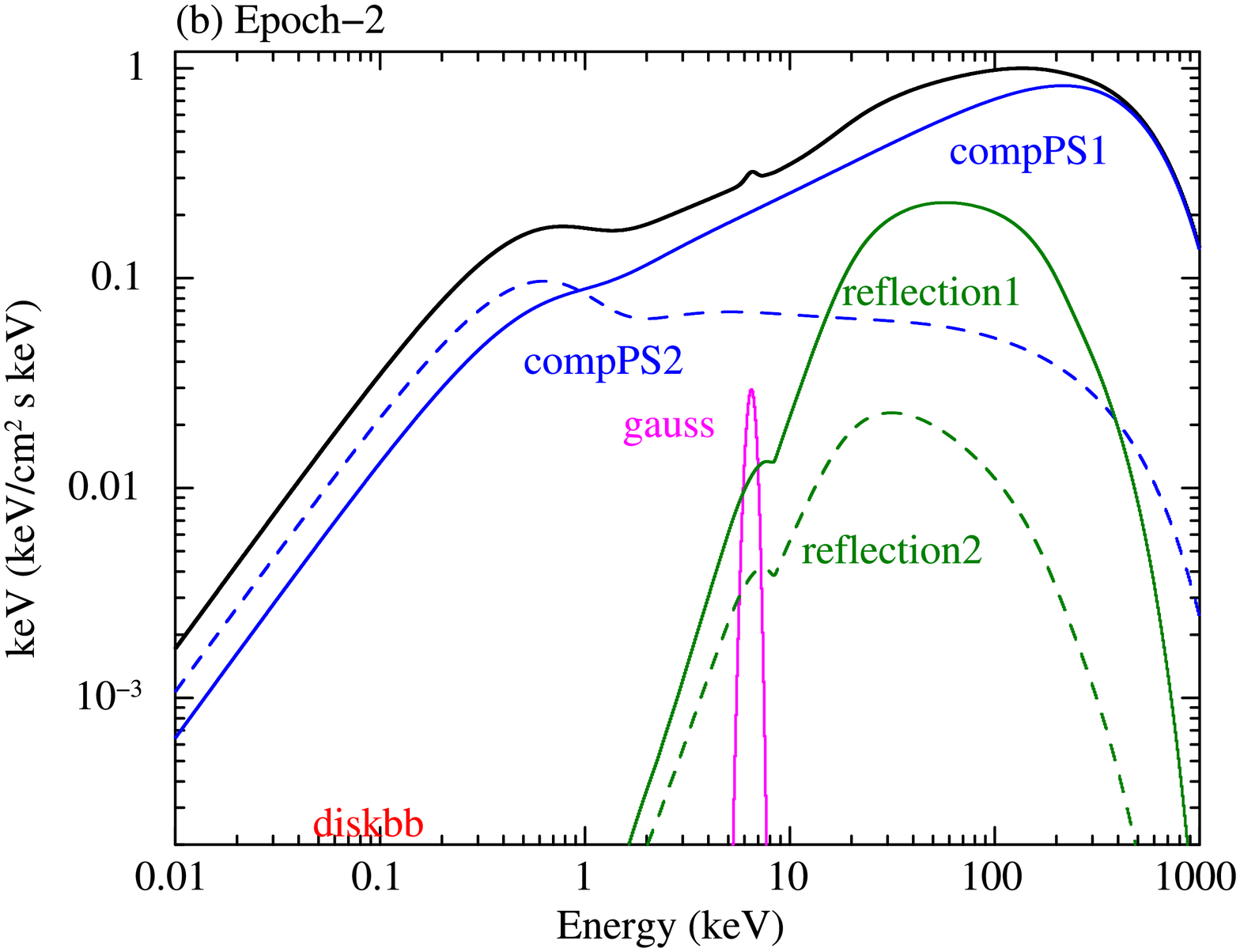}
\FigureFile(70mm,){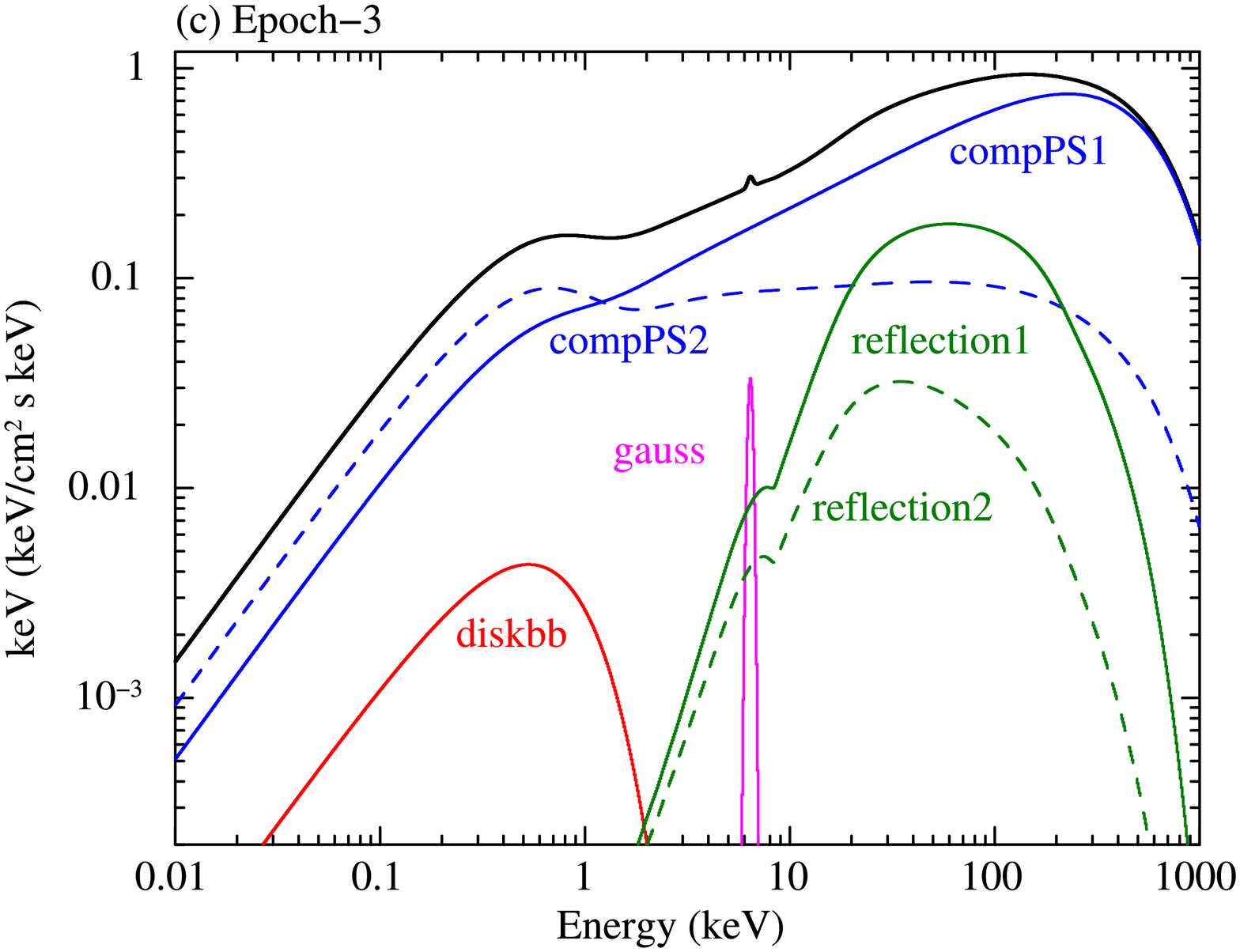}
\end{center}
\caption{The best-fit double {\tt compPS} model of each epoch in the $\nu F_{\nu}$
form, corrected for the interstellar absorption. Each component
is separately plotted. \label{model_comp}}
\end{figure*}

\begin{table*}
\caption{Parameters of the Double Compton Model\footnotemark[$*$] for 
Epoch-1, -2, and -3\label{tbl-2}}
\begin{center}
\begin{tabular}{lllll}
\hline
Component & Parameter & Epoch-1 & Epoch-2 & Epoch-3 \\
\hline
wabs & $N_{\rm H}$ $(10^{21} {\rm cm}^{-2})$ & $ 5.35 \pm 0.03$ & $5.52^{+0.05}_{-0.03}$ & $ 5.46\pm0.03$ \\ \hline
diskbb & $T_{\rm in}$ (keV) & $0.223 \pm 0.003$ & $ 0.220^{+0.002}_{-0.02}$ & $ 0.225^{+0.002}_{-0.003}$ \\
& normalization\footnotemark[$\S$] & $<$120 & $<$255 & $ 250^{+120}_{-110}$ \\
& photon flux\footnotemark[$\|$] & $1.59\times 10^{-3}$ & $1.51\times 10^{-3}$ & $4.09\times 10^{-2}$ \\ \hline

compPS\footnotemark[$\dagger$] & $T_{\rm e}$ (keV) & $176^{+4}_{-6}$ & $171^{+6}_{-5}$ & $177^{+5}_{-6}$ \\
(harder component) & (y-parameter\footnotemark[$\#$] & $1.55^{+0.04}_{-0.01}$ & $1.2 \pm 0.4$ & $1.24^{+0.17}_{-0.01}$) \\ 
 & $\tau$ & $ 1.12^{+0.07}_{-0.03}$ & $ 0.9 \pm 0.3 $ & $ 0.90^{+0.04}_{-0.03}$ \\
& photon flux\footnotemark[$\|$] & 0.283 & 0.679 & 0.547 \\ \hline
compPS\footnotemark[$\dagger$] & (y-parameter\footnotemark[$\#$] & $0.63\pm0.01$ & $0.40 \pm 0.02$  & $ 0.51^{+0.02}_{-0.01}$) \\ 
(softer component) & $\tau$ & $ 0.46 \pm0.02$ & $0.30^{+0.02}_{-0.01} $ & $ 0.37\pm0.02$ \\
& photon flux\footnotemark[$\|$] & 1.10 & 0.889 & 0.798 \\ \hline
reflection\footnotemark[$\ddagger$] & $\xi$ (erg cm s$^{-1}$) & $ 0.474^{+1.561}_{-0.470}$ & $<1.71$ & $0.481^{+1.767}_{-0.477}$ \\
 & $\Omega/(2\pi)$ & $ 0.46 \pm 0.01$ & $0.48^{+0.02}_{-0.01}$ & $ 0.43^{+0.02}_{-0.01}$ \\
Gaussian & $E_{\rm cen}$ (keV) & $ 6.41^{+0.04}_{-0.06}$ & $ 6.44\pm 0.06$ & $ 6.40^{+0.04}_{-0.03}$ \\
 & EW (eV) & $ 96^{+9}_{-11}$ & $ 99^{+10}_{-13}$ & $ 58^{+7}_{-8}$\\
 & $\sigma$ (keV) & $ 0.35^{+0.10}_{-0.03}$ & $0.38^{+0.11}_{-0.06}$ & $0.18^{+0.04}_{-0.03}$ \\ \hline
& $R_{\rm in}^{\rm cont}$ (km)\footnotemark[$**$] & $117\pm2$ & $128_{-2}^{+22}$ & $115\pm2$ \\ 
& $\chi^2/{\rm d.o.f.}$ & $996/946$ & $873/903$ & $940/869$ \\ \hline
Unabsorbed flux & 0.5--2 keV (erg cm$^{-2}$ s$^{-1}$) & $ 3.5\times10^{-10}$ & $ 3.8\times10^{-10}$ & $ 3.5\times10^{-10}$ \\
 & 2--10 keV (erg cm$^{-2}$ s$^{-1}$) & $ 6.4\times10^{-10}$ & $ 6.6\times10^{-10}$ & $ 6.2\times10^{-10}$ \\
 & 10--300 keV (erg cm$^{-2}$ s$^{-1}$) & $ 4.4\times10^{-9}$ & $ 4.3\times10^{-9}$  & $ 4.0\times10^{-9}$ \\ 
\hline
\multicolumn{5}{@{}l@{}}{\hbox to 0pt{\parbox{180mm}{\footnotesize
Notes. The errors are 90\% confidence level for a single parameter.
\par\noindent
\footnotemark[$*$] The model is expressed as ${\tt wabs}\times(
{\tt compPS}+{\tt compPS}+{\tt diskbb}+{\tt gauss})$. 
\par\noindent
\footnotemark[$\dagger$] We set gmin $=-1$ (thermal electrons only) and 
geom $=4$ (spherical geometry). The seed photons for 
\par
Comptonization is the MCD emission from the disk (i.e., $T_{\rm bb}$ in {\tt compPS} 
is linked as $-T_{\rm in}$ in {\tt diskbb}).
\par\noindent
\footnotemark[$\ddagger$] We assume the solar abundances for iron and heavy
elements. The temperature of the reflector is fixed at $10^6$ K. 
\par
The reflection parameters of the two {\tt compPS} components are linked 
each other, with $R_{\rm in}$, $R_{\rm out}$, $\beta$, and $i$ fixed 
\par
at the best-fit values from the iron-K line fit summarized in Table~\ref{tbl-3}.
\par\noindent
\footnotemark[$\S$] $((r_{\rm in}/{\rm km})/(D/10 {\rm kpc}))^2 \cos{\theta}$, 
where $r_{\rm in}$ and $\theta$ are the inner disk radius and the disk inclination, respectively.
\par\noindent
\footnotemark[$\|$] Unabsorbed photon flux in units of photons cm$^{-2}$
s$^{-1}$ in the 0.01--100 keV band based on the best-fit parameters.
\par\noindent
\footnotemark[$\#$] Compton y-parameter calculated as $4\tau k T_{\rm e}/(m_{\rm e}c^2)$.
\par\noindent
\footnotemark[$**$] Estimated innermost disk radius derived from the flux of direct MCD and Comptonization 
components, 
\par
assuming a distance of 8 kpc and an inclination of 50$^\circ$. Those have been corrected for 
the inner boundary condition
\par
and color/effective temperature ratio (see Section 4.2).
}\hss}}
\end{tabular}
\end{center}
\end{table*}

To best constrain the origin of the X-ray emission, the broad-band
spectra of the XIS and HXD are simultaneously analyzed. We use the
data in the energy range of 0.5--8.0 keV for XIS-1, 1.0--8.0 keV for
XIS-0 plus -3, 15--50 keV for PIN, and 50--310 keV for GSO. 
To avoid the calibration uncertainties around the instrumental Si-K
and Au-M edges in the responses, the spectra in the 1.7--1.9 keV and
2.1--2.3 keV energy band are excluded in the fit. The cross
normalization between HXD and XIS-0$+$XIS-3 spectra is fixed at 1.18
based on the calibration result using the Crab Nebula data
\footnotemark[4], while that between XIS-1 and XIS-0$+$XIS-3 is set
free.

As shown in Figure \ref{spec123}, the spectra of the three
observations exhibit similar shape to one another, indicating that the
physical state of the source remain the same in the three epochs. We
find that the overall continuum in the 2--100 keV band can be roughly
approximated by a power law of a photon index of $\sim$1.5 in each
epoch. This confirms that GX 339--4 was the low/hard state throughout
our observations.

Following the previous studies of BHBs in the low/hard state with Suzaku, 
we adopt a model from a standard accretion disk and its
thermal Comptonization by a hot corona as the basic description of the
broad band continuum. The disk emission is assumed to be the
multi-color blackbody radiation modeled by {\tt diskbb}
\citep{mit84}. For the Comptonization component, we employ the {\tt
compPS} model \citep{pou96}, where the Comptonized spectrum is
computed from the electron temperature, the optical depth of
scattering, and the energy distribution of incident photons. We
consider only thermal electrons by setting gmin$=-1$. We assume that
the Comptonization cloud is spherical (i.e., geom$=4$) and that seed
photons are entirely originated in the standard disk ($T_{\rm
bb}=-T_{\rm in}$ in the XSPEC terminology).

The {\tt compPS} code is able to take into account a reflection
component from the disk as calculated by \citet{mag95}. Its 
parameters are the solid angle of the reflector visible from the source
($\Omega/2\pi$), ionization parameter ($\xi$), inclination angle
($i$), and temperature, which we fix at 10$^6$ K. 
This component is further relativistically
blurred with the {\tt diskline} profile \citep{fab89}, by assuming a
radial dependence of emissivity as $\propto R^{\beta}$ between the
inner and outer disk radius ($R_{\rm in}$ and $R_{\rm out}$). We fix
$R_{\rm in}=10 R_{\rm g}$, $R_{\rm out}= 10000R_{\rm g}$,
$i=50^{\circ}$, and $\beta = -2.3$, based on the results from the
analysis of the XIS spectra in the 3--8 keV band
(Section~4.3). Although Compton broadening is not taken into account
in the reflection model, the effect is negligible in our case because
the ionization parameter is found to be very low ($\xi \lesssim 2$,
Section~4.1) and estimated temperature does not exceed $\sim 10^6$ K
as assumed. The iron K fluorescent line at $\sim 6.4$ keV is simply
fitted with a Gaussian at this stage, whose detailed modeling does not
affect the analysis of the broad band continuum. Combining all the
components described above, the model is expressed as ${\tt
wabs}\times ({\tt compPS}+{\tt diskbb}+{\tt gauss})$ on the XSPEC
terminology.

As the first trial, we perform a spectral fit without reflection
components. The result is far from acceptable, yielding
$\chi^2/\nu = 1219/950$ (Epoch-1), $1160/907$ (Epoch-2), and
$1195/873$ (Epoch-3), with residuals shown in the second panels of
Figure~\ref{fit_comp}. As noticed, there remains an absorption
edge-like feature above $\sim 7$ keV in the XIS spectra and a hump
structure at $\sim 30$ keV in the PIN data. They are strong evidence
for the reflection component, as reported in previous broad band
observations of GX 339--4 in the same state (e.g., \cite{ued94}).

We find that inclusion of the reflection component in the model
greatly improve the fit, although the fit goodness is not yet
satisfactory, 
$\chi^2/\nu = 1161/948$, $1083/905$, and $1136/871$ for Epoch-1,
-2, and -3, respectively. There is still a discrepancy
between the data and model in the 15--50 keV region, as shown in the
third panels of Figure \ref{fit_comp}. 
We confirm that this is not caused by any systematic
uncertainties in the response or background of PIN, based on the
calibration using the Crab Nebula and comparison of the background
model with the dark-Earth occultation data in our observations.
Similar residuals are also found from other BHBs in the low/hard state
when fitted with the same model (GRO J1655--40 by \cite{tak08} and Cyg
X-1 by \cite{mak08}). 
A possible reason for the discrepancy is that the corona cannot be
described by a constant temperature and optical depth, and therefore
we add another {\tt compPS} component. This ''double {\tt compPS}''
model was successfully applied to the low/hard state spectra of BHBs,
such as Cyg X-1 \citep{mak08,gie97} and GRO J1655--40
\citep{tak08}. We tie all the parameters of the additional {\tt
compPS} component to those of the other {\tt compPS} component except
for the optical depth $\tau$ and normalization.

The best-fit model and residuals of fit using the double {\tt compPS}
model are displayed in the top and bottom panels of Figure \ref{fit_comp},
respectively. In Figure \ref{model_comp}, the best-fit model spectrum in
the $\nu F_{\nu}$ form (where $F_{\rm \nu}$ is the energy flux) is
plotted, with contribution of each component, corrected for the
interstellar absorption. The best-fit values are summarized in 
Table~\ref{tbl-2}.
The fit is significantly improved, $\chi^2/\nu =996/946$,
$873/903$, $940/869$, compared with the single {\tt compPS} model, yielding
F-test probabilities of $4\times10^{-32}$, $7\times10^{-43}$, and $1\times10^{-36}$ 
for Epoch-1, -2, and -3, respectively.
We find that the contribution of the direct MCD component ({\tt
diskbb}) is very small ($< 2\%$ in flux at 0.5 keV), and only its
upper limit is obtained in Epochs~1 and 2. The absorption column
density is determined as $N_{\rm H} \approx 5.4 \times 10^{21}$
cm$^{-2}$, which is consistent with previous works by \citet{men97}
and \citet{kon00}.

\subsection{Estimate of the Innermost Disk Radius}

Assuming that the number of photons emitted from the disk is conserved
in the Comptonization process, we can estimate the innermost 
radius of the disk from the following equation \citep{kub04},
\begin{eqnarray}
F^p_{\rm disk}+F^p_{\rm thc}2 \cos i =& 0.0165 \left[ \frac{r_{\rm in}^2\cos i}{(D/10\mbox{ kpc})^2}\right] 
\left( \frac{T_{\rm in}}{1\mbox{ keV}} \right)^3 \nonumber \\
 & \mbox{ photons } {\rm s}^{-1} \mbox{ }{\rm cm}^{-2}. \label{eq1}
\end{eqnarray}
where $F^p_{\rm disk}$ and $F^p_{\rm thc}$ are the unabsorbed 0.01--100 keV
photon flux from the disk and thermal Compton component, respectively.
Here it is assumed that the Comptonizied emission is isotropic 
and there are few photons that are scattered back to the disk.
Table~\ref{tbl-2} list the 
values of $F^p_{\rm disk}$ and 
$F^p_{\rm thc}$ (for the two Comptonization components) in each epoch;
the direct disk flux is found to be less than 3\% of the total Comptonized one.
Assuming $i = 50^\circ$ (Section~4.2),
we derive the innermost disk radius for Epoch-1, -2 and -3 to be
$r_{\rm in}=(98 \pm 2) D_{8}$, $(107_{-2}^{+19}) D_{8}$, $(97\pm 2)
D_{8}$ km, respectively, where $D_{8}$ is the distance to GX 339--4 in
the unit of 8 kpc. Here the errors reflect the uncertainties in
$T_{\rm in}$. Multiplying this radius by the correction factor,
1.19, for the boundary condition and ratio of the color/effective
temperature of the multi-color disk (see \cite{kub98}), yields an
actual innermost radius of $R_{\rm in}^{\rm cont}=(117 \pm 2)
D_{8}$, $(128_{-2}^{+22}) D_{8}$, $(115\pm 2) D_{8}$ km for the three
epochs.

\subsection{Iron-K Line Fitting}

To best constrain the disk geometry from the iron-K emission line
profile, we fit the XIS spectra in the 3--8 keV band with the
relativistic disk line model {\tt diskline} \citep{fab89} with the
reflection continuum self-consistently. Here we consider both
K$\alpha$ and K$\beta$ lines;
although the contribution of the K$\beta$ line is generally weak and
has been neglected in most of previous work, 
we find that it affects the final results.
Since the parameters can be only weakly constrained from the
spectra in each epoch due to the limited photon statistics, we
combined the spectra of the three epochs separately for XIS-0$+$XIS-3
and for XIS-1 to discuss the time-averaged properties over the three
observations. This is justified because the continuum fit to the
individual broad band spectra gives similar (if not the same) results
to one another, and the estimated inner disk radii ($R_{\rm in}^{\rm cont}$) 
stay constant within 10\% level as shown in the previous
subsection, which is much smaller than the statistical error we will
obtain from the line profile analysis.
As a continuum, we adopt the results from double {\tt compPS} fit
using the broad-band XIS$+$HXD spectra averaged over the three epochs,
and replace the {\tt gauss} model with {\tt diskline} for the iron
line. In the fit of the XIS spectra, the normalization and optical
depth of the hard Comtonization component, the most dominant one in
the 3--8 keV band, are treated as free parameters. Since the continuum
shape is hard to constrain within the limited band coverage, they are
allowed to float only within their 90\% errors obtained in the
XIS$+$HXD fit, and the other continuum parameters are fixed at the
best-fit values.

\begin{figure}
\begin{center}
\FigureFile(80mm,){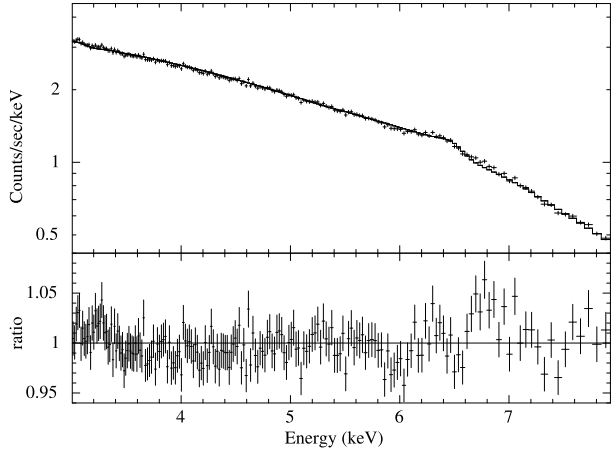}
\end{center}
\caption{Iron K-line fit to the 3--8 keV XIS-0$+$XIS-3 spectra in 2009 March with the
{\tt diskline} model for a fixed emissivity index of $\beta=-3$.
The ratio between the data and best-fit model are shown in the lower panel.\label{beta-3}}
\end{figure}

\begin{figure}
\FigureFile(80mm,){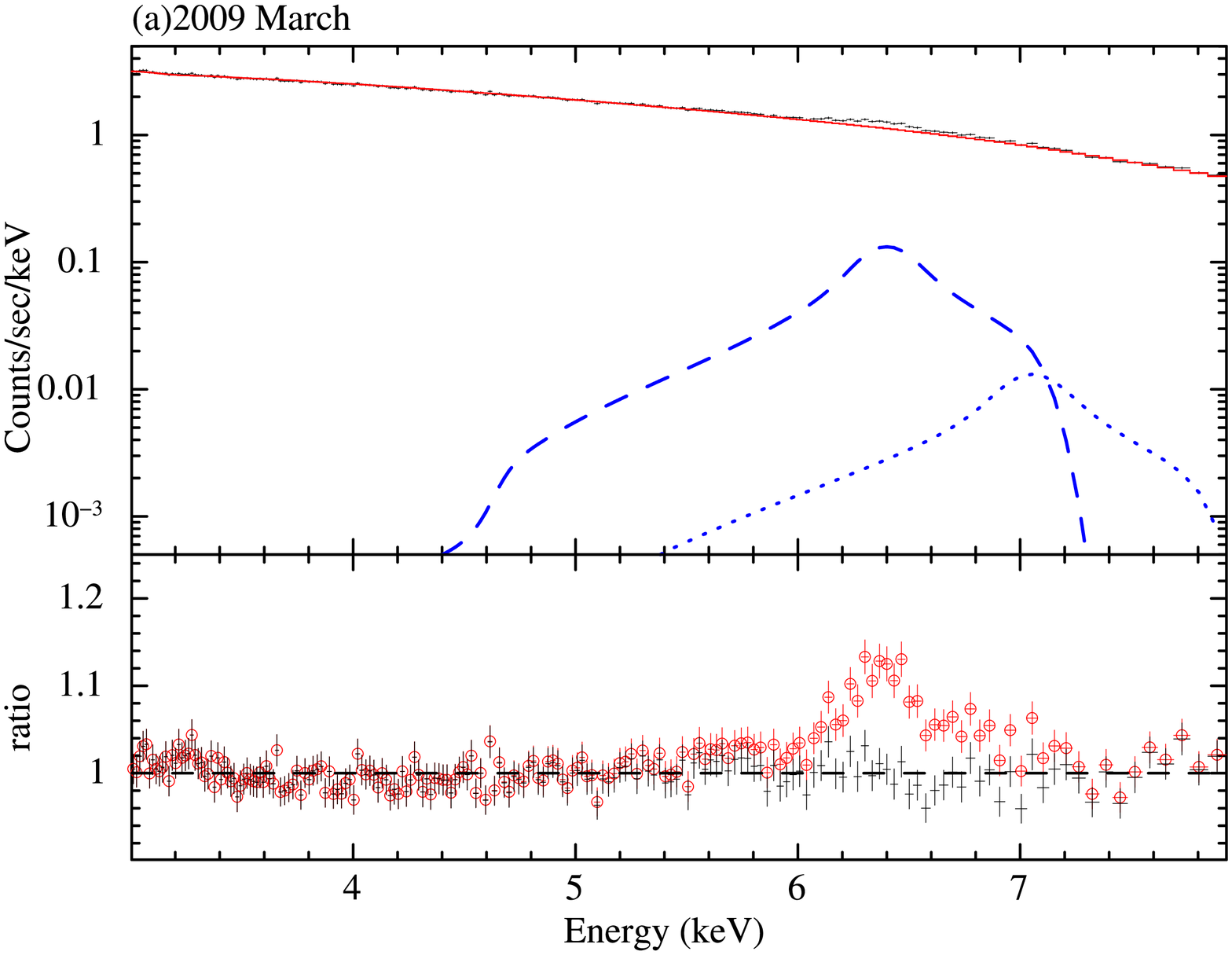}
\FigureFile(80mm,){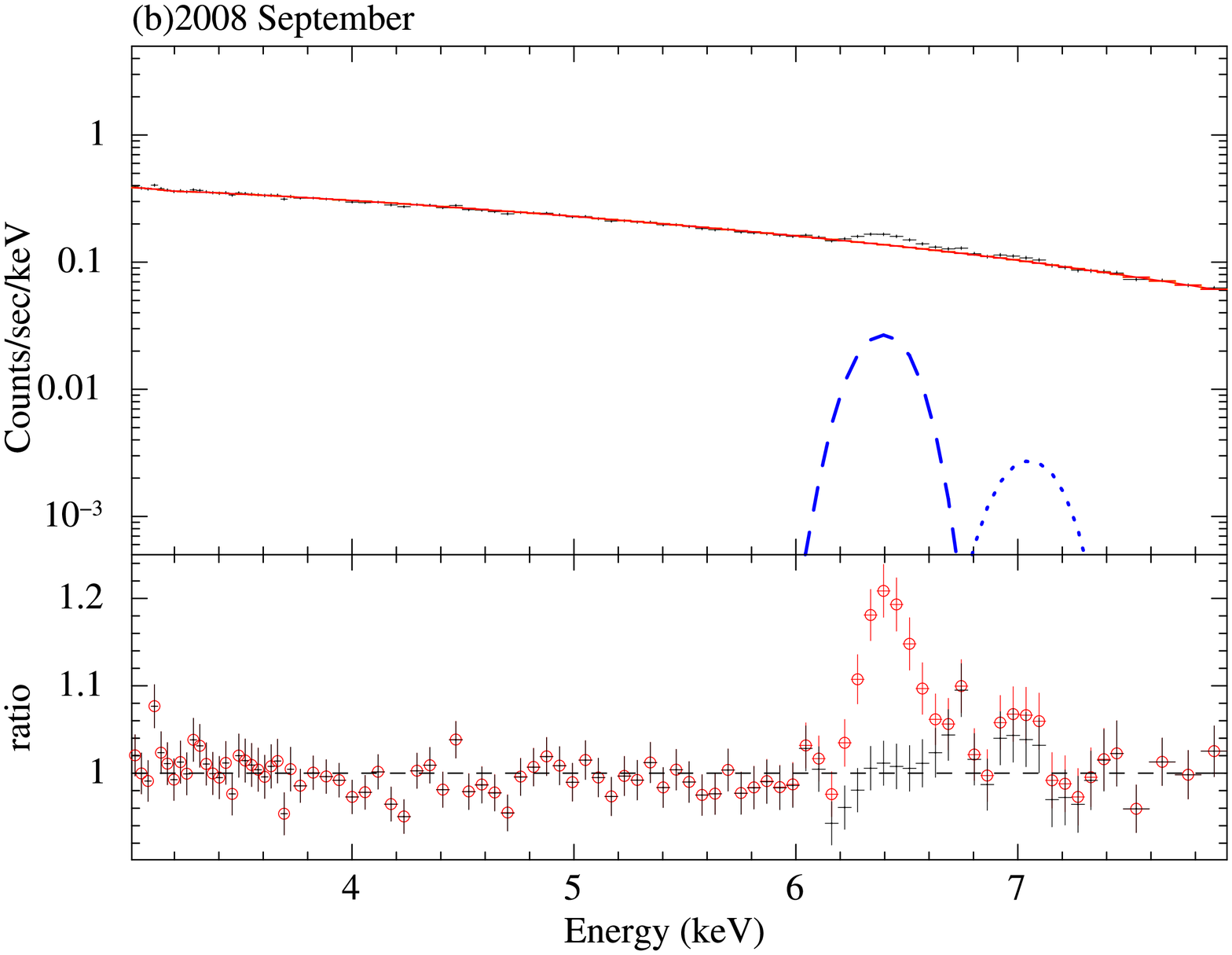}
\caption{
Fitting results of the 3--8 keV XIS-0$+$XIS-3 spectra with the {\tt
diskline} model, obtained from the observations in 2009 March (our
data, brighter state) and that in 2008 September \citep{tom09}. The
top panel plots the data (black, with error bars) and the best-fit
models of the iron-K$\alpha$ (blue, dashed) and -K$\beta$ lines (blue,
dotted) and of the continuum (red, solid). The ratio between the 
data and best-fit model are shown in the lower panels for
the 2009 and 2008 data, respectively; the red ones (open circles)
correspond to the case when the iron-K line is excluded from the
model. For clarity the XIS-1 data are not plotted.\label{fe_line} }
\end{figure}

\begin{table}
\caption{Parameters of the Iron-K Line Fit\footnotemark[$*$]\label{tbl-3}}
\begin{center}
\begin{tabular}{lll}
\hline 
Component & Parameter & Value \\ 
\hline
diskline (K$\alpha$) & $E_{\rm cen}$ (keV) & 6.4 (fixed) \\
 & EW (eV) & $87^{+10}_{-11}$ \\
diskline (K$\beta$) & $E_{\rm cen}$ (keV) & 7.06 (fixed) \\
 & Emissivity law $\beta$ & $-2.3\pm0.1$ \\
 & $i$ (deg) & $46\pm8$ \\
 & $R_{\rm in}$ ($R_{\rm g}$) & $13.3^{+6.4}_{-6.0}$ \\
 & intensity ratio (K$\beta /$K$\alpha$) & 0.13 (fixed) \\ \hline 
& $\chi^2/{\rm d.o.f.}$ & $273/271$ \\ \hline
\multicolumn{3}{@{}l@{}}{\hbox to 0pt{\parbox{85mm}{\footnotesize
Notes. The errors are 90\% confidence level for a single parameter.
\par\noindent
\footnotemark[$*$] The continuum model is ${\tt wabs}\times({\tt diskbb}+{\tt
compPS}+{\tt compPS})$.
}\hss}}
\end{tabular}
\end{center}
\end{table}

\begin{figure}
\begin{center}
\FigureFile(75mm,){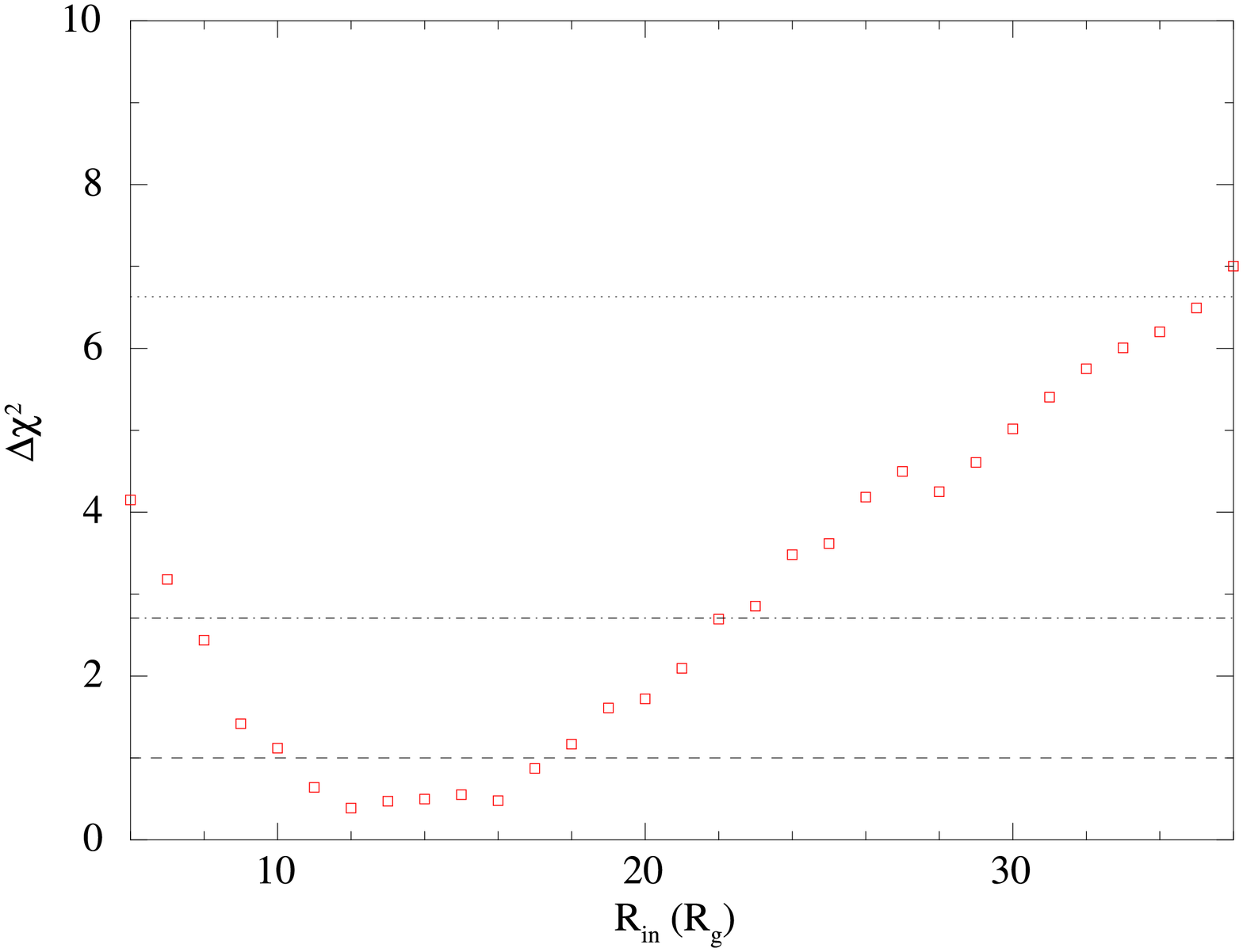}
\FigureFile(75mm,){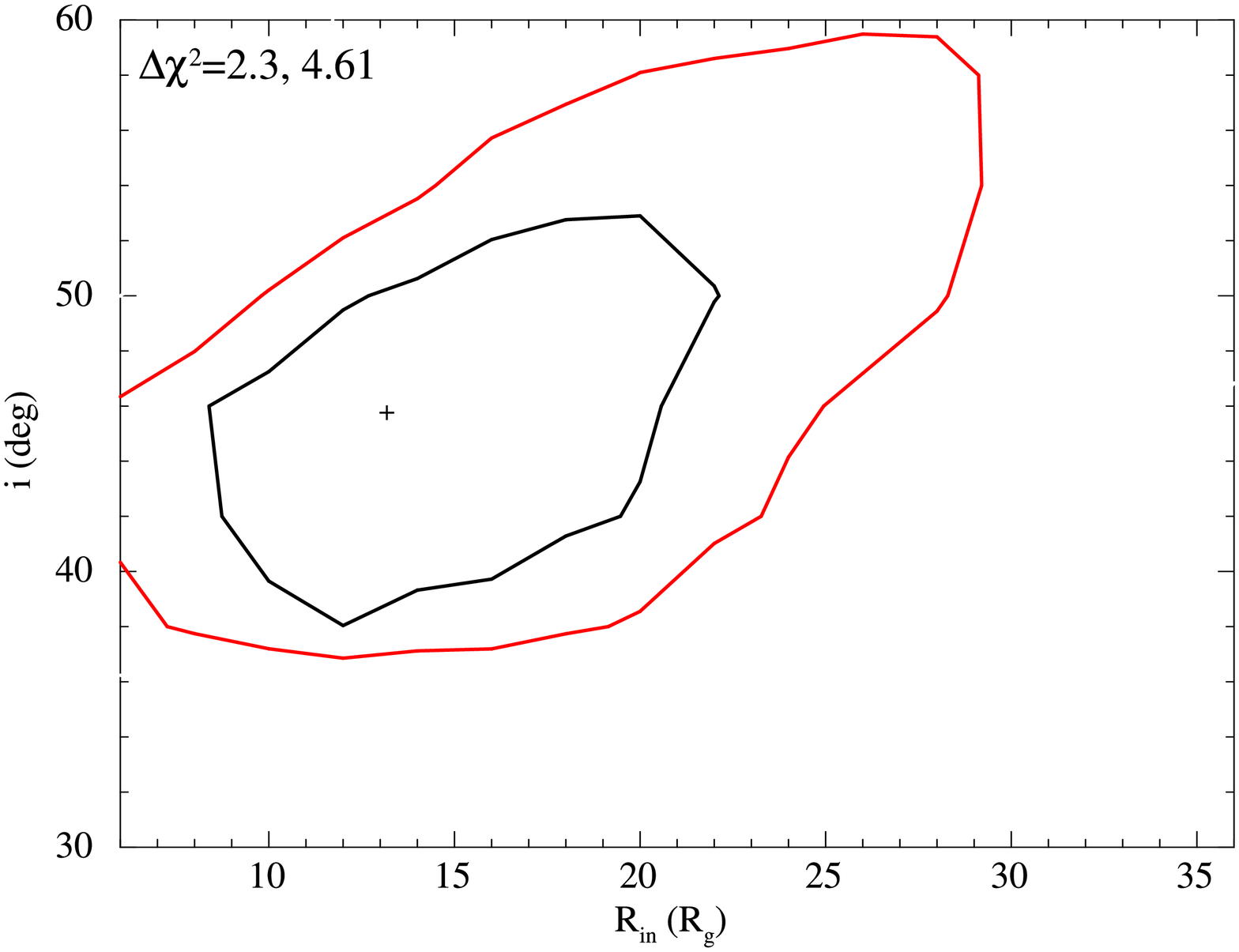}
\FigureFile(75mm,){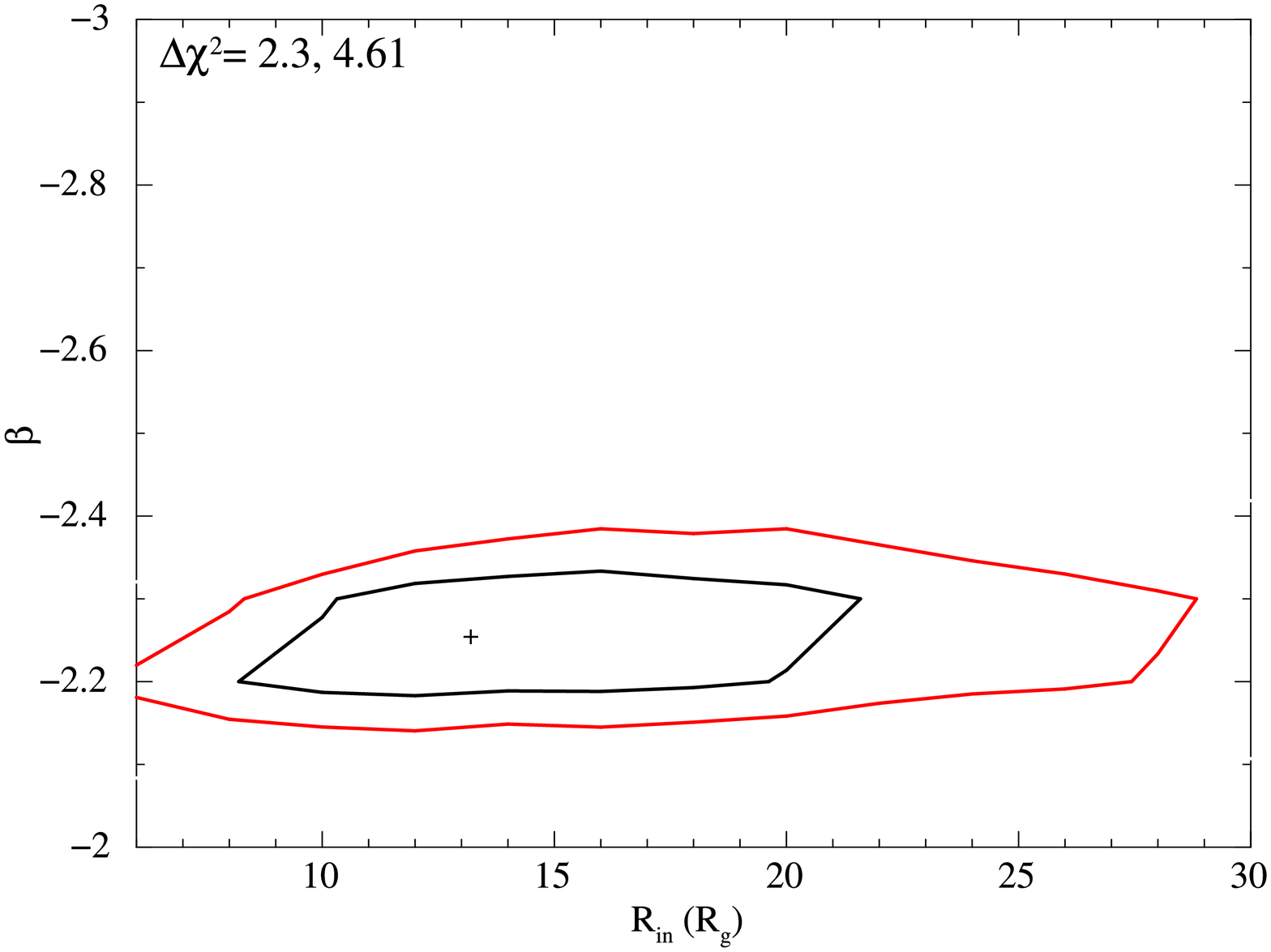}
\end{center}
\caption{
{\it Top}: Delta chi-squared values plotted against $R_{\rm in}$
obtained from the {\tt diskline} fit. The dashed, dash-dotted, and dotted
lines represent 68\%, 90\%, and 99\% confidence limits for a single parameter, 
respectively. {\it Middle}:
Confidence contour between the inner radius $R_{\rm in}$ and 
inclination $i$ at 90\% (red, outer) and 68\% (black, inner) 
confidence limits for two parameters.
{\it Lower}: Confidence contour between the inner radius $R_{\rm in}$ and 
emissivity index $\beta$ at 90\% (red, outer) and 68\% (black, inner) 
confidence limits for two parameters.\label{steppar}
}
\end{figure}

The {\tt diskline} model has five parameters, the inner and
outer disk radius ($R_{\rm in}$ and $R_{\rm out}$), inclination angle
($i$), power law index of emissivity $\beta$, and intrinsic line
center energy ($E_{\rm cen}$). We fix $R_{\rm out} = 10000 R_{\rm g}$,
which is difficult to constrain from the data.
The line center energies are fixed at 6.40 keV (K$\alpha$) and 7.06
keV (K$\beta$) to make them consistent with the low ionization
parameter obtained from the reflection continuum. We also fix the
intensity ratio between the K$\beta$ and K$\alpha$ at 0.13, 
a theoretical value for the iron K-fluorescence process \citep{kaa93}. The
rest are left as free parameters. Because the reflection continuum
must also be blurred in the same manner as the emission line, we tie
$R_{\rm in}$, $R_{\rm out}$, and $\beta$ in the {\tt diskline} model
with those of the reflection component included in the {\tt compPS}
model. The result on the inclination can be compared with the
independent constraints presented in Figure~\ref{incl_D}. When the
emissivity index is fixed at $\beta = -3$, we find that the fit is
not acceptable ($\chi^2/{\rm d.o.f.}= 323/272$), showing a significant
discrepancy between the data and model in both sides of the iron line
(see Figure~\ref{beta-3}). In addition, the best-fit inclination angle for
$\beta=-3$ is found to be $\approx 20^\circ$, which is not reasonable since
the black hole mass exceeds $100 M_{\solar}$ according to
Figure~\ref{incl_D}, unless the inner accretion disk is heavily
warped with respect to the binary plane. Thus, we vary $\beta$ in the
range $-3 \leq \beta \leq-2$ as a free parameter.

The result of the fit is plotted in Figure~\ref{fe_line}, and the
best-fit parameters are listed in Table~\ref{tbl-3}. We also show
the increment of $\chi^2$ in terms of $R_{\rm in}$, and confidence
contour between $R_{\rm in}$ and $i$, and that between $R_{\rm in}$
and $\beta$ in Figures~\ref{steppar} (a), (b), and (c), respectively.  The
inclination angle now becomes $i=46\pm 8$ degrees, which is
reasonable in terms of the black hole mass (Figure~\ref{incl_D}).
The resulting inner radius is $R_{\rm in}= 13.3^{+6.4}_{-6.0}
R_{\rm g}$, suggesting that the standard disk is extended close to the
ISCO of a non-rotating black hole but is likely truncated at 90\%
confidence level (see Figure~\ref{steppar}).

To compare with $R_{\rm in}^{\rm cont}$, the unit of the innermost
disk radius derived through the iron-K line fit must be converted to
km; $R_{\rm in}=13.3^{+6.4}_{-6.0} R_{\rm g}$ corresponds to $R_{\rm
in}=200^{+100}_{-90} M_{10}$ km, where $M_{10}$ is the black hole
mass in the unit of 10 $M_{\solar}$.  Thus, the results from the
continuum and iron-K line profile analysis become consistent if the BH
mass is the range of (4--16)$M_{\solar}$, assuming the distance of 8
kpc.

\subsection{Comparison with 2008 Suzaku data}

\citet{tom09} observed GX 339--4 in the fainter low/hard state
using Suzaku, from 2008 September 24 22:37:25 to 27 02:19:24
(UT) with a net exposure of 100 ksec for the XIS and 80 ksec for the
HXD (ObsID: 403067010). The 1--100 keV flux of the source was $2.4
\times 10^{-10}$ erg cm$^{-2}$ sec$^{-1}$ in this period, which is
approximately 14 times smaller than that of our 2009 data. As
described in Section~4.1, the best-fit $R_{\rm in}$ value in the 2009
data is $R_{\rm in}=13.3^{+6.4}_{-6.0}$, apparently a factor of $>$8
smaller than the 2008 result by \citet{tom09}. We note, however, that
the derived $R_{\rm in}$ is coupled with both inclination and
emissivity index. To eliminate the uncertainties, we re-analyze the
data of the 2008 observation. The data reduction is made according to
the standard procedure from the cleaned event files of processing
version 2.2.11.22. We extract the XIS time-averaged spectra in the 
0.5--10.0 keV band. The XIS-0 and XIS-3 spectra are co-added together 
with their responses. 
 
We analyze the spectra with the model composed of an absorbed
power law plus two iron line (diskline) components for K$\alpha$ and
K${\beta}$ line. For the analysis, we use the XIS-0$+$XIS-3 and XIS-1
spectra in the 3--8 keV range. We fix the inclination angle at
$i=50^\circ$ and the emissivity index at $\beta=-2.3$ according to our
results of the Suzaku 2009 data. The quality of the fit is found to be
acceptable, $\chi^2/\nu = 444/402$, with the best-fit photon index of
$\Gamma = 1.61 \pm 0.01$ and column density of $N_{\rm H}=5.31 \pm
0.07 \times 10^{21}$. The XIS spectra, best-fit model, and their ratio
are plotted in Figure~\ref{fe_line}.  Table~\ref{tbl-tom} summarizes
the best-fit parameters. It is seen that the iron-K lines in the 2008
data is significantly narrower than those observed in 2009. The
resulting inner disk radius is found to be $R_{\rm
in}=190^{+710}_{-90}$, which is $\approx$14 times larger (at least 5 times
larger within the statistical errors) than that in our observations.

\begin{table}
\caption{Parameters of the Iron-K Line Fit to the \citet{tom09} Data\footnotemark[$*$]\label{tbl-tom}}
\begin{center}
\begin{tabular}{lll}
\hline
Component & Parameter & Value \\
\hline
diskline (K$\alpha$) & $E_{\rm cen}$ (keV) & $6.4$ (fixed) \\
 & EW (eV) & $77^{+18}_{-13}$ \\
diskline (K$\beta$) & $E_{\rm cen}$ (keV) & 7.06 (fixed) \\
 & Emissivity law $\beta$ & $-2.3$ (fixed) \\
 & $i$ (deg) & $50$ (fixed) \\
 & $R_{\rm in}$ ($R_{\rm g}$) & $190^{+710}_{-90}$ \\ 
 & intensity ratio (K$\beta /$K$\alpha$) & 0.13 (fixed) \\ \hline 
& $\chi^2/{\rm d.o.f.}$ & $444/402$ \\
\hline
\multicolumn{3}{@{}l@{}}{\hbox to 0pt{\parbox{85mm}{\footnotesize
Notes. The errors are 90\% confidence level for a single parameter.
\par\noindent
\footnotemark[$*$] The continuum model is ${\tt wabs}\times{\tt
powerlaw}$.
}\hss}}
\end{tabular}
\end{center}
\end{table}

\section{Near-Infrared \hspace{-0.1mm}Observations \hspace{-0.1mm}and \hspace{-0.1mm}Data \hspace{-0.1mm}Analysis}

We carried out near-infrared photometric observations of GX 339--4 on
11 nights over a period from 2009 Feb 27 to 2009 March 23, using the
SIRIUS camera \citep{nag03} on the 1.4 m IRSF telescope at the South
African Astronomical Observatory (SAAO). SIRIUS is a {\it JHK}$_{\rm
s}$-simultaneous imaging camera, which has three 1024 $\times$ 1024
HAWAII arrays and covers a field of view of $7'.7 \times 7'.7$ with a
scale of $0''.45$ pixel$^{-1}$. The log of observations is given in
Table~\ref{tbl_irsf}. The typical seeing (FWHM) during the
observations was $\sim1''.4$ (3 pixels) in the {\it J} band.

\begin{figure}
\begin{center}
\FigureFile(75mm,){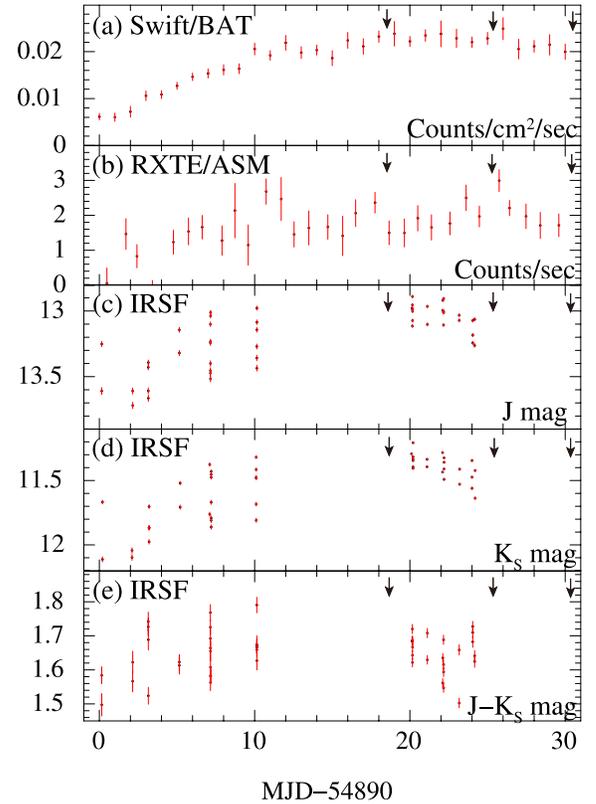}
\end{center}
\caption{
X-ray (15--50 keV and 2--12 keV) and IR ($J$ and $K_{\rm S}$ band) light
curves and $J-K_{\rm s}$ magnitudes of GX~339--4 obtained with Swift/BAT, 
RXTE/ASM, and IRSF/SIRIUS in 2009 March. The
abscissa represents the date of MJD$-54890$, where 1 corresponds to
2009 March 1. The IR magnitudes are calibrated based on the Two
Microns All Sky Survey (2MASS) Point Source Catalog, by using stars in
the field of view. The arrows in each panel show the epochs of our
Suzaku observations.
\label{lc_irX}}
\end{figure}

After a standard data reduction procedure for near-infrared array
images using the IRAF analysis package (dark subtraction, flat-field
correction and sky subtraction), 10 dithered images were combined to
reject bad pixels and improve the S/N ratio. Hence, each magnitude in
Figure~\ref{lc_irX} was determined from 50 or 100 sec
integration. Aperture photometry was made with an aperture of 12-pixel
radii. Photometry with smaller apertures gave similar results. We used
the 2MASS data of about 150 stars in the field for magnitude
calibration. The 1-day averaged magnitudes in the {\it JHK}$_{\rm s}$ 
band are listed in Table~\ref{tbl_irsf}.

\begin{table*}
\caption{The Log of Near-infrared Observations\label{tbl_irsf}}
\begin{center}
\begin{tabular}{cccccc}
\hline
Date\footnotemark[$*$] & Number of Observations\footnotemark[$\dagger$] & 
Integration Time per Frame (sec) & \multicolumn{3}{c}{Averaged Magnitude} \\
 & & & $J$ & $H$ & $K_{\rm S}$ \\
\hline
Feb 28 & 2 & 5 & $13.43 \pm 0.03$ & $12.60 \pm 0.02$ & $11.89 \pm 0.03$ \\
Mar 2 & 2 & 5 & $13.66 \pm 0.03$ & $12.80 \pm 0.02$ & $12.07 \pm 0.03$ \\
Mar 3 & 4 & 5 & $13.52 \pm 0.04$ & $12.67 \pm 0.03$ & $11.85 \pm 0.03$ \\
Mar 5 & 2 & 5 & $13.23 \pm 0.03$ & $12.40 \pm 0.02$ & $11.61 \pm 0.02$ \\
Mar 7 & 9 & 5 & $13.27 \pm 0.06$ & $12.40 \pm 0.04$ & $11.63 \pm 0.05$ \\
Mar 10 & 6 & 5 & $13.21 \pm 0.05$ & $12.30 \pm 0.03$ & $11.53 \pm 0.03$ \\
Mar 20 & 7 & 10 & $13.00 \pm 0.03$ & $12.12 \pm 0.02$ & $11.33 \pm 0.02$ \\
Mar 21 & 2 & 10 & $13.03 \pm 0.02$ & $12.16 \pm 0.01$ & $11.36 \pm 0.01$ \\
Mar 22 & 6 & 10 & $12.99 \pm 0.03$ & $12.11 \pm 0.02$ & $11.38 \pm 0.02$ \\
Mar 23 & 2 & 10 & $13.05 \pm 0.02$ & $12.17 \pm 0.01$ & $11.47 \pm 0.01$ \\
Mar 24 & 5 & 10 & $13.16 \pm 0.03$ & $12.30 \pm 0.02$ & $11.49 \pm 0.02$ \\ \hline
\multicolumn{3}{@{}l@{}}{\hbox to 0pt{\parbox{85mm}{\footnotesize
\footnotemark[$*$] All observations were made in 2009.
\par\noindent
\footnotemark[$\dagger$] We carried out 10 ditherings at each time.
}\hss}}
\end{tabular}
\end{center}
\end{table*}

\begin{figure*}
\begin{minipage}{0.5\hsize}
\begin{center}
\FigureFile(80mm,){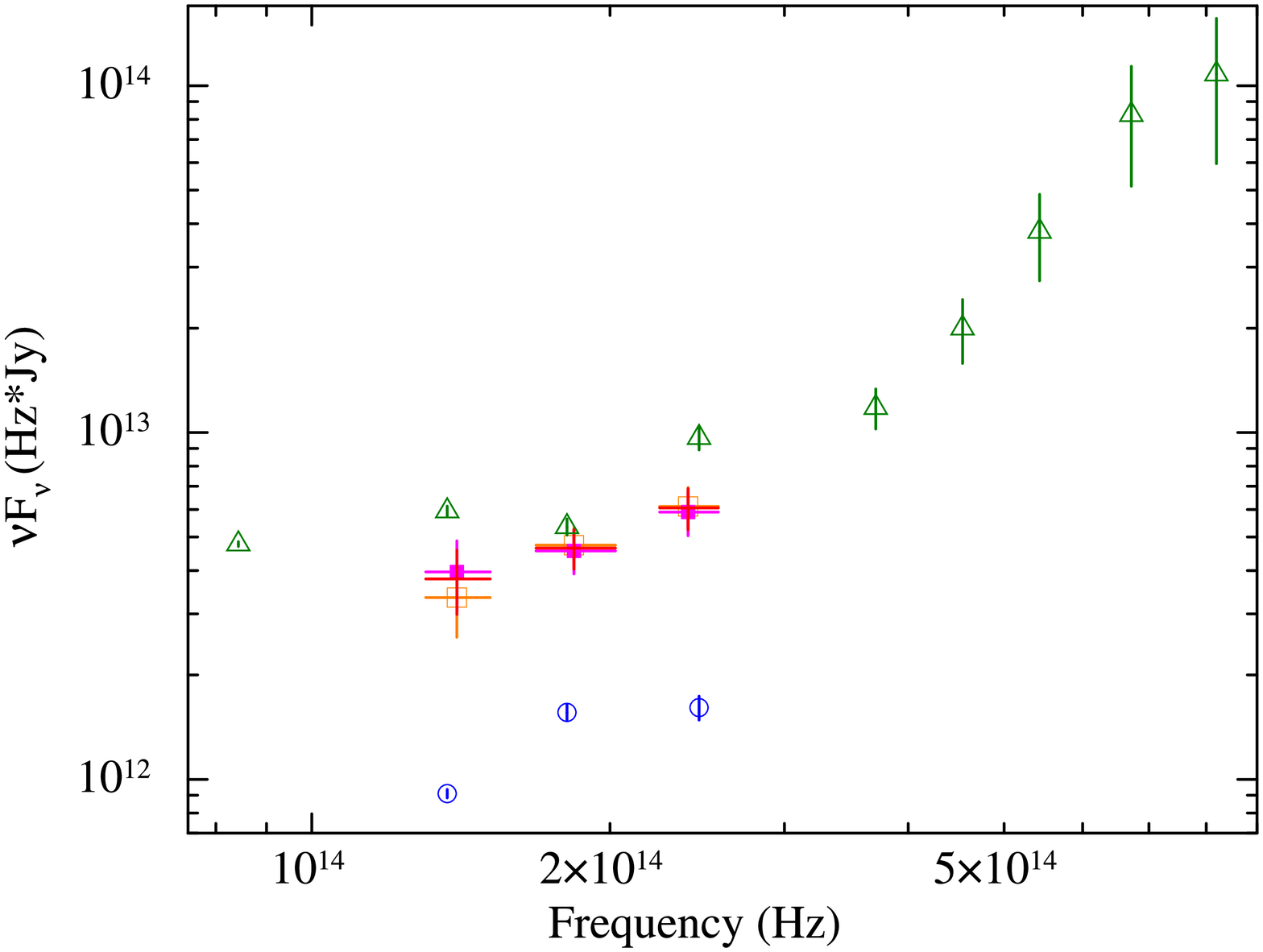}
\end{center}
\end{minipage}
\begin{minipage}{0.5\hsize}
\begin{center}
\FigureFile(80mm,){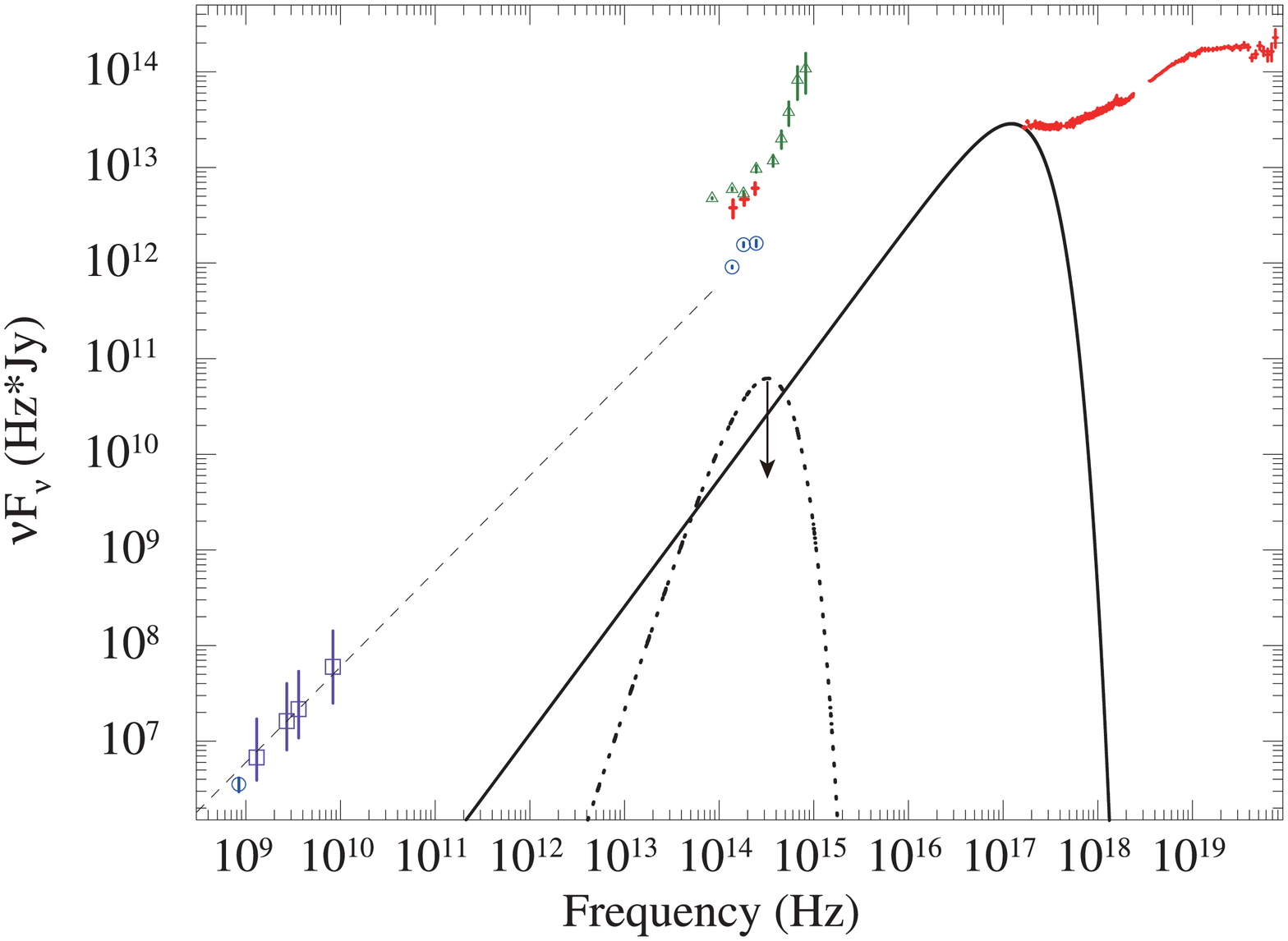}
\end{center}
\end{minipage}
\caption{Multiwavelength SED of GX 339--4 in the $\nu F_\nu$ form,
corrected for the interstellar absorption/extinction. In the left
panel, the data from our IRSF/SIRIUS observations, those
obtained in 1981 and 1997 \citep{cor02} are shown in red, 
green (open triangle), and
blue (open circle), respectively. The IRSF fluxes averaged over MJD
54910--54915 are plotted in red, that with the reddest $J-K_{\rm s}$
color are in pink (filled square), and with the bluest color in orange (open square).
In the right panel, the Suzaku spectra and the radio data in
1992--1999 \citep{fen01} and 1997 \citep{cor02} are displayed in red,
purple (open square), and blue (open circle), respectively, in
addition to the same data as the left panel (without the
reddest/bluest $J-K_{\rm s}$ data).
The solid line shows the estimated contribution of intrinsic
multicolor disk emission, including the Compton scattered photons.
The dotted line represents the upper limit of the contribution from
the companion star, assuming a blackbody with an effective temperature
of $4000$~K \citep{zdz04}. The dashed line shows a power law with an
index of 1.0 ($\alpha=0.0$), fitted to the \citet{fen01} results, for
illustrative purpose.
\label{SED}}
\end{figure*}

The $J$ ($1.25$ $\mu$m) and $K_{\rm s}$ ($2.14$ $\mu$m) light curves 
with the $J-K_{\rm s}$ color are plotted in Figure~\ref{lc_irX},
together with those of Swift/BAT in the 15--50 keV band and RXTE/ASM 
in the 2--12 keV band. As noticed, the daily averaged $K_{\rm s}$ band 
flux is correlated positively with the hard X-ray flux. Significant 
variation in the $J-K_{\rm s}$ color by $<$0.3 magnitudes is 
observed on time scale of 100 sec, although we find no clear correlation 
between the color and magnitude. The $H$ ($1.63$ $\mu$m) band light curve 
is not plotted because they basically follow the same trend as the $J$ 
and $K_{\rm s}$ light curves.

The left panel of Figure~\ref{SED} shows the SED of GX 339--4 at the
near-infrared to optical wavelengths in the $\nu F_{\nu}$
form. Utilizing the IRSF data taken between 2009 March 20 to 24,
when the Swift/BAT flux was nearly constant around the Suzaku
observation epoch, we plot the averaged photometry of the
$J$-$H$-$K_{\rm s}$ bands, the spectrum with the reddest $J-K_{\rm s}$
color, and that with the bluest color. For comparison, the near
infrared to optical data acquired in 1981 and 1997 in the low/hard
state are also plotted \citep[and references therein]{cor02}. In the
right panel, a combined SED of GX 339--4 over the radio to X-ray bands
is shown, although they are not simultaneous except for the IRSF and
Suzaku data, which are quasi-simultaneous. The interstellar
absorption is corrected in the X-ray spectrum, and all the near
infrared data are deredenned by assuming $A_V=3.7$ \citep{cor02}. Both
X-ray and infrared fluxes in our observations are between those
obtained in 1981 and 1997, which differ by a factor of 4. By fitting
the $J$-$H$-$K_{\rm s}$ spectra by a power law with $F_{\nu} \propto
\nu^{\alpha}$, we obtain $\alpha = -0.10\pm0.08$ (average), which
varied between $\alpha=-0.20\pm0.15$ (the reddest data) and
$\alpha=+0.08\pm0.10$ (bluest).

\section{Discussion}

\subsection{Suzaku Results and Inner Disk Structure}

With Suzaku, we have obtained the best quality simultaneous,
broad band X-ray spectra ever obtained from GX 339--4 in the low/hard
state, covering the 0.5--310 keV band. We demonstrate that the
spectra in the three observations
can be well described by thermal Comptonization of disk
photons dominating the flux of the entire band, a direct MCD
component, and a reflection component with an iron-K emission line. 
The disk photon flux is very small ($\lesssim$3\% of the
Comptonized one), indicating that the inner part of the optically
thick disk is almost fully covered by the corona. The strength of the
reflection component (in terms of the solid angle, $\Omega/2\pi
\approx 0.45$) and the ionization parameter of the reflector
($\xi \lesssim 2$) are consistent with the result from a
previous Ginga study in the similar intensity state (in 1989
September) analyzed with a power law continuum \citep{ued94}, but are
now determined with much smaller uncertainties thanks to the better
energy resolution and wider coverage of Suzaku.

We find that in each epoch, two thermal Comptonization components
(``double CompPS'' model), with the same electron temperature ($T_{\rm
e} \approx 175$ keV) but different optical depths ($\tau \approx$ 0.4
and 1), gives a reasonable description of the incident continuum
spectrum.  This indicates that the Comptonizing corona can be regarded
isothermal but has a more complex geometry than that assumed in a
single zone, spherical model. In fact, the corona may have a more
disk-like geometry and/or inhomogeneous structure, where the optical
depths of Comptonization as measured from the input source (i.e., the
standard disk) is not spatially constant. Similar conclusions have
been obtained from Cyg X-1 \citep{mak08} and GRO J1655--40
\citep{tak08} in the low/hard state with Suzaku. Thus, we
suggest that such physical conditions of the corona may be a common
feature in the low/hard state of BHBs.
Although the overall shape of the broad-band spectra of GX~339--4
look similar during our observations, the optical depths and relative
fraction of the two Comptonization components are found to be slightly
different among the three epochs while the electron temperature
remains constant at $\approx$175 keV. This suggests that
the geometry of the corona varies on time scale of several days in the
low/hard state even at an almost same ($<5$\%) luminosity level.
We note that the double CompPS fit still leaves
small, systematic discrepancies between the data and model in the PIN
$\lesssim 30$ keV region. This is not surprising since the change of
physical parameters of the corona should be continuous in reality,
while the double CompPS model gives only an approximation. Application
of more complex (and realistic) Comptonization models is left as
future work.

The analysis of the iron-K line profile has enabled us to tightly
constrain the inclination angle and innermost radius of the optically
thick disk. We obtain $i \sim 50$ degrees, which is converted to the
black hole mass of 4--16 $M_{\solar}$ for the allowed range of the
mass function of this binary (see Section~2). The estimated mass range
looks more reasonable than huge masses ($>100$ $M_{\solar}$) calculated
from $i=18$ degrees, if the disk is not strongly warped, as claimed by
\citet{mil06} and \citet{rei08} based on their detection of a broad
iron-K line. The reason for this large discrepancy is unclear, but it
should be noted that results on a very broad iron-K line often
strongly depend on the continuum modeling and may be partially
affected by pile-up of the CCD data \citep{yam09,don10}.

The emissivity index obtained from the empirical diskline fit also
carries important information for understanding the disk structure. In
our state, a part of the iron-K line emitted from the innermost radii
should be subject to Comptonization as well, under the situation that
the hot corona almost fully covers the inner disk. This causes
significant broadening of the scattered line component, with an
averaged fractional energy gain of $\Delta E/E \sim 4kT_{\rm e}/m c^2
\sim 1$, which makes it almost undetectable as a line. This would
partially explain the observed flatter slope of the emissivity,
$\beta=-2.3$, than the simplest expectation of $\beta=-3$ when a
perfectly plain disk is irradiated from a point source with a finite
scale height. Another reason for the flat slope is that the index will
approach to $\beta=-2$ at outer radii as the scale height of the
irradiating source becomes negligible compared with that of the disk.

As mentioned in Section~4.3, the innermost radius obtained from the
diskline fit, in units of $R_{\rm g}$, are consistent with that
independently estimated from the MCD continuum, in units of km, within
the uncertainties of the mass function and distance.  Comparing our
$R_{\rm in}$ value with those obtained in previous observations, we
can study how the inner edge of the accretion disk evolves during the
low/hard state as a function of luminosity. The 1--100 keV flux during
our observations is approximately 14 times brighter than that in the
2008 Suzaku observation \citep{tom09}, which was $2.4 \times 10^{-10}$
erg cm$^{-2}$ sec$^{-1}$.  As described in Section~4.3, the 
$R_{\rm in}$ value ($R_{\rm in}=13.3^{+6.4}_{-6.0}$) is
at least 5 times smaller within the statistical errors
than that obtained in our re-analysis
of 2008 data by adopting the common inclination
($i=50^\circ$) and emissivity index ($\beta=-2.3$). These results
indicate that the inner edge moved inward as the luminosity increased
from 2008 to 2009.

\begin{table*}
\caption{Summary of Previous Results of Iron-K Line Fit Using a Relativistic Disk-line Model.\label{tbl-rin}}
\begin{center}
\begin{tabular}{cccccccl}
\hline
State & Satellite & $L_{1-100 {\rm keV}}$ 
& $R_{\rm in}$ & $i$ & $\beta$ 
& $E_{\rm cen}$ & Ref. \\
 &  & (\% $L_{\rm Edd}$)
& ($R_{\rm g}$) & (deg) &  
& (keV) & \\
\hline
low/hard & Suzaku$+$RXTE & 0.14 & $>65$ & 18 (fixed) 
& ($-2$) -- ($-3$) & $6.47^{+0.04}_{-0.03}$ & [9] \\
&Suzaku & 0.14 & $190^{+710}_{-90}$ & 50 (fixed) 
& $-2.3$ (fixed) & $6.4$, $7.06$\footnotemark[$\S$] (fixed) & [1]\footnotemark[$\|$] \\
&Swift$+$RXTE & 0.46 & $2.9^{+2.1}_{-0.7}$ & 20 (fixed) & $-3.1\pm0.4$ 
& $6.7^{+0.4}_{-0.3}$ & [8] \\
&Swift$+$RXTE & 1.33 & $3.6^{+1.4}_{-1.0}$ &  20 (fixed) & $-3.2^{+0.6}_{-0.5}$ 
& $6.9^{+0.2}_{-0.5}$ & [8] \\
&Suzaku & 2.0 &  $13.3^{+6.4}_{-6.0}$ & $46\pm8$ 
& $-2.3\pm0.1 $ & $6.4$, $7.06$\footnotemark[$\S$] (fixed) & [1] \\
& XMM-Newton$+$RXTE & 3.25 & $ 4.0\pm 0.5$ 
& $20^{+5}_{-15}$ & $-3.0$ & $6.8\pm0.1$ & [5] \\
&XMM-Newton & 3.25&$2.8\pm0.1$ & $10^{+2}_{-0, {\rm pegged}}$ 
& $-3.23^{+0.04}_{-0.05}$ & $6.97^{+0.01}_{-0.06}$ & [7]\footnotemark[$\#$] \\
&XMM-Newton & 3.25& $6.33^{+1.04}_{-0.08}$ & $22^{+1}_{-3}$ & $-3.4\pm2$ & 
$7_{-0.02}$ & [2]\footnotemark[$\#$] \\
&XMM-Newton & 3.25&  $24^{+1}_{-4}$ & 60 (fixed) &$ -3$ (fixed) 
& $6.4^{+0.01}$ & [2]\footnotemark[$\#$] \\ \hline
very high & XMM-Newton & 12 & $2.1^{+0.2}_{-0.1}$ & $11^{+5}_{-1}$ 
& $-5.5^{+0.1}_{-0.5}$ & $6.97_{-0.20}$ & [4] \\
&XMM-Newton & 12& $1.91^{+0.02}_{-0.01}$ & $18.2^{+0.3}_{-0.5}$ 
& $-6.82^{+0.04}_{-0.03}$ & $6.97_{-0.01}$ & [7]\footnotemark[$**$] \\
&XMM-Newton$+$RXTE & 12& $35^{+25}_{-3}$ &  60 (fixed)
& & & [3]\footnotemark[$**$] \\
& Suzaku & (28)\footnotemark[$*$] & $\sim1$ \footnotemark[$\dagger$]
& $18\pm1$ & $-3.0\pm0.1$ & & [6]\\
& Suzaku & (28)\footnotemark[$*$] & 8.2($^{+5.8}_{-3.2}$)\footnotemark[$\ddagger$] 
& $33^{+12, {\rm pegged}}_{-8{, \rm pegged}} $ & $-3$ (fixed) 
& 6.68$^{+0.41}_{-0.47}$  & [10]\footnotemark[$\dagger\dagger$] \\ \hline
\multicolumn{8}{@{}l@{}}{\hbox to 0pt{\parbox{180mm}{\footnotesize
Notes. The errors are 90\% confidence level for a single parameter unless otherwise noted.
\par\noindent
\footnotemark[$*$] 0.5--200 keV Luminosity.
\par\noindent
\footnotemark[$\dagger$] {\tt CDID} reflection model convolved with {\tt kerrdisk} was used.
\par\noindent
\footnotemark[$\ddagger$] 1-sigma confidence limit.
\par\noindent
\footnotemark[$\S$] Iron K$\beta$ line.
\par\noindent
\footnotemark[$\|$] Re-analysis of the \citet{tom09} data.
\par\noindent
\footnotemark[$\#$] Re-analysis of the \citet{mil06} data.
\par\noindent
\footnotemark[$**$] Re-analysis of the \citet{mil04} data.
\par\noindent
\footnotemark[$\dagger\dagger$] Re-analysis of the \citet{mil08} data.
\par\noindent
References. [1] this paper; [2] \citet{don10}; 
[3] \citet{kol11}; [4] \citet{mil04}; [5] \citet{mil06}; 
\par
[6] \citet{mil08}; [7] \citet{rei08}; [8] \citet{tom08}; 
[9] \citet{tom09}; [10] \citet{yam09} 
}\hss}}
\end{tabular}
\end{center}
\end{table*}

Table~\ref{tbl-rin} summarizes major previous results on the innermost radius
$R_{\rm in}$ of the standard disk of GX~339--4 in different X-ray
luminosities and states, all derived from analysis of the iron-K
emission line with a relativistic disk-line model. We also list other
line parameters including the inclination and emissivity index, with
which $R_{\rm in}$ could be coupled. The original results obtained by
\citet{mil06} in the low/hard state and \citet{mil08} in the very high
state have been recently re-examined by \citet{don10} and
\citet{yam09}, respectively, and it is pointed out that both data were
significantly affected by pile-up. The re-analysis show that the data
favor larger values of $R_{\rm in}$ in both cases, although the new
constraints become weaker due to poorer photon statistics.

\begin{figure}
\begin{center}
\FigureFile(80mm,){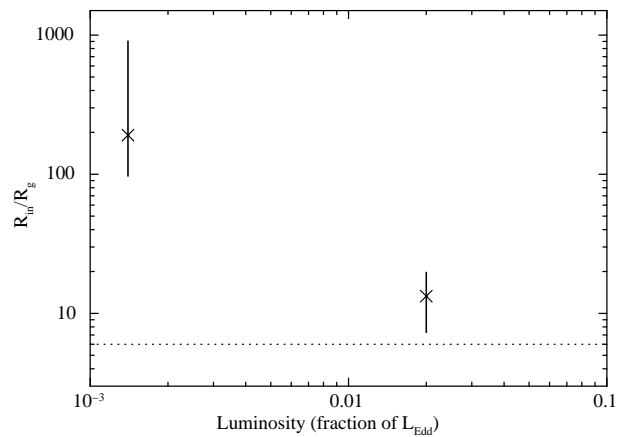}
\caption{
Estimated innermost radii of the standard disk of GX 339--4 derived
from our results from the 2008 (left) and 2009 (right) Suzaku data with
an inclination of $i = 50^\circ$ and an emissivity index of $\beta=-2.3$,
plotted against fraction of Eddington luminosity ($L_{\rm X}$/$L_{\rm Edd}$).
\label{fig4_tom}}
\end{center}
\end{figure}

Figure~\ref{fig4_tom}
plots $R_{\rm in}$ with its 90\% statistical error for a single
parameter, as a function of
Eddingtion fraction $L_{\rm X}/L_{\rm Edd}$ ($D$ = 8 kpc and black
hole mass of $10 M_{\solar}$ are assumed). 
This figure indicates that the standard accretion disk of GX~339--4
evolves inward as the luminosity increases, with a significant change 
in the range of $\sim 0.001 < L_{\rm X}/L_{\rm Edd} < \sim 0.02$ when
the source remains in the low/hard state, as concluded by
\citet{tom09}. Our result in the bright low/hard state in 2009 is
consistent with that of \citet{don10} obtained at a similar luminosity
if $R_{\rm in} \approx 20 R_{\rm g}$. This suggests that at $L_{\rm
X}/L_{\rm Edd} = 0.02$ the standard disk is likely truncated before
reaching to 6$R_{\rm g}$, the ISCO for a non-spinning black hole,
although our Suzaku result alone cannot rule out the possibility
that the inner edge lies below 6$R_{\rm g}$ at 99\% confidence. If GX
339--4 has an extremely-rotating black hole as claimed by
\citet{mil08}, our 2009 result then would be strong evidence that the
disk does not extend to the ISCO in the bright low/hard state. Further
observations with similar or better energy resolution not limited by
the photon statistics and systematic errors caused by pile-up would be
necessary to further clarify the right location of the inner edge of
the disk in the bright ($\sim 0.01 L_{\rm Edd}$) low/hard state.

\subsection{IRSF Results and Jet Energetics}

We examine the origin of the multiwavelength SED of GX 339--4 in the
low/hard state with different emission components, utilizing our
quasi-simultaneous near infrared and X-ray spectra combined with
previous data. In the right panel of Figure~\ref{SED}, we plot the
estimated contribution of the ``intrinsic'' MCD component including
photons that are Compton-scattered in the corona, through the photon
number conservation, assuming that the disk is extending to infinity
(for illustrative purpose) with the temperature proportional to
$r^{-3/4}$, where $r$ is the radius from the black hole. The results
are based on the double CompPS fit, which gives the innermost disk
temperature of 0.22 keV. We also plot the upper limit of the
contribution from the companion star, assuming a blackbody of
4000~K. Here we refer to the $r$-band magnitude of $>21.4$ by \citet{zdz04},
which is then corrected for extinction corresponding to $A_V = 3.7$.
As noticed from the figure, the blackbody component of the multi-color
disk and companion star contributes only $\lesssim 0.5$\% and
$\lesssim 1$\% of the total flux in the near-infrared range,
respectively. In the radio band, a slightly inverted power law
spectrum is observed with an energy index of $\alpha \approx$0.1--0.2
\citep{cor00}, which is interpreted as an optically thick synchrotron
emission from the compact jets, a common feature observed from BHBs in
the low/hard state.

Our averaged IRSF spectrum shows a somewhat smaller energy index,
$\alpha \approx -0.1$, than the radio spectrum; the corresponding
$J-K_{\rm s}$ color is similar to that of the 1981 observations
compiled by \citet{cor02}. As shown by them, this color is consistent
with a superposition of two different components with $\alpha=-0.6$
and $\alpha=2.1$, which can be explained by the optically {\it thin}
synchrotron radiation from the jets and the reprocessed, thermal
emission by X-ray irradiation from outer parts of the disk,
respectively (see also \cite{mar03,gan10}). It is thus indicated that
the transition point from optically thick to thin regime of the
synchrotron emission, defined as $\nu_{\rm t}$, must appear at
frequencies just below or around the near-infrared band, $\sim
10^{14}$ Hz. Note that this conclusion does not match the SSC model
discussed by \citet{cor09} for the low/hard state of GX~339--4 where
the H band corresponds to an optically {\it thick} emission in our
luminosity range ($L/L_{\rm Edd} \sim 0.01$ for $M_{\rm BH} = 10
M_{\solar}$ and $D=8$ kpc). The rapid variability observed in the
optical band (e.g., \cite{gan08}) suggests that the power law
component of synchrotron radiation extends there, although it is not
clear at which frequency the spectrum breaks due to radiative
cooling. In our picture, it should be below the X-ray band, since the
X-ray flux must be dominated by thermal Comptonization that can well
account for the Suzaku spectra.

The observed synchrotron luminosity and the limit for $\nu_{\rm t}$
can be used to constrain the physical parameters of the jet base, from
which most of the observed near infrared luminosity is emitted
according to the compact jet model \citep{bla79}. For
simplicity, here we consider a single zone model representing this
region.  We neglect any effects of relativistic beaming from the bulk
motion of the jet, since the Doppler factor is estimated to be $\delta
\approx$1.03 for an inclination of $50^\circ$ and an assumed intrinsic
jet velocity of 0.9$c$. Let us assume that the electron number density
at the Lorentz factor $\gamma$ is given as $n(\gamma) = A \gamma^{-p}$
(for $p>2$) in the range between $\gamma_{\rm min} < \gamma <
\gamma_{\rm max}$. The synchrotron luminosity density in an optical thin
region is expressed as (see e.g., \cite{ryb79})
\begin{eqnarray*}
\nu L_\nu = & \frac{3^{\frac{p+2}{2}}}{2\pi (p+1)}
\; \Gamma \left(\frac{p}{4}-\frac{1}{12} \right) \Gamma \left(\frac{p}{4}+\frac{19}{12} \right) \\
& \; \times V A u_B \sigma_{\rm T} c \left(\frac{\nu}{\nu_B} \right)^{\frac{3-p}{2}} \left( \sin \theta \right)^{\frac{p+1}{2}},
\end{eqnarray*}
where $V$ is the volume ($V = 4\pi R^3/3$ for a sphere with
a radius of $R$), $u_B \equiv B^2/(8 \pi)$ is the 
energy density of the magnetic field, $\sigma_{\rm T}$ is the cross
section of Thomson scattering, $\nu_B \equiv e B / (2\pi m c)$ is 
the gyro frequency ($e$ and $m$ are the electronic charge and the electron mass, 
respectively), $\Gamma(y)$ is the gamma function of argument y,
and $\theta$ is the pitch angle, which is the angle between magnetic field
and electron velocity. By assuming equipartition of energy density between the
electrons' kinetic energy and the magnetic field, that is $ A = (p-2)u_{\rm
B}/(mc^2) $ with $\gamma_{\rm min} = 1$, we can estimate the strength
of the magnetic field from an observed luminosity for given
$R$ and $p$. Further, using an strict form of 
the synchrotron absorption coefficient,
\begin{eqnarray*}
\alpha_{\nu}= & \frac{(p-2)}{16 \pi^2} 3^{\frac{p+3}{2}}
\; \Gamma \left(\frac{3p+2}{12} \right) \Gamma \left(\frac{3p+22}{12} \right) \\
& \; \times u_{B}^{2} \frac{\sigma_{\rm T}}{m^2 c}
\nu_B^{-3} \left(\frac{\nu}{\nu_B}\right)^{-\frac{(p+4)}{2}}
\left( \sin \theta \right)^{\frac{p+2}{2}},
\end{eqnarray*}
the frequency $\nu_{\rm t}$ below which the synchrotron emission
becomes optically thick for a region size of $R$ can be determined
from the condition $\alpha_{\nu} R = 1$.  Taking $\nu L_\nu \approx
3\times10^{35}$ erg s$^{-1}$ at $\nu = 1.4\times10^{14}$ Hz (for
$D$=8 kpc), $p=2.2$ (corresponding to the spectral index
$\alpha=(1-p)/2=-0.6$), and $\nu_{\rm t} \sim 10^{14}$ Hz, we estimate
$B \approx 5\times10^4$ G and $R \approx 6\times10^8$ cm, which corresponds
to $\sim 4\times10^2 R_{\rm g}$.
More precisely, dependence of $B$ on $R$ and that of $\nu_{\rm t}$ on $B$ and $R$ 
are given as
$$
B \approx
5\times10^4 \;
\left(\frac{\nu L_\nu (K)}{3\times10^{35}\; {\rm erg\;s}^{-1}}\right)^{\frac{2}{p+5}} \;
\left(\frac{R}{6\times10^8\; {\rm cm}}\right)^{-\frac{6}{p+5}} {\rm \;\;G}
$$
and
$$
\nu_{\rm t} \approx 
1\times10^{14} \;
\left(\frac{B}{5\times10^4\; {\rm G}}\right)^{\frac{p+6}{p+4}} \;
\left(\frac{R}{6\times10^8\; {\rm cm}}\right)^{\frac{2}{p+4}} {\rm \;\;Hz},
$$
respectively, where the terms of the pitch angle are averaged (i.e., 
integrated over solid angle and normalized by $4\pi$). 
For this $B$ value, 
the corresponding energy density of the magnetic field of $8.4\times10^7$
erg cm$^{-3}$ and the Lorentz factor of electrons emitting the 
$\nu=1.4 \times 10^{14}$ Hz photons is $\gamma \approx 40$.
Our results hold within a factor of 2 when we assume
``particle-dominant'' energy balance with $u_{\rm e} \sim 10 u_B$, as
observed from jets of blazars (e.g., \cite{ino96}).
The magnetic-field strength in the jet base is consistent with
the \citet{gia05} result of $B=10^5$ G obtained for XTE J1118+480 
in the low/hard state from modelling of the multi-wavelengths SED.

Based on these parameters, we roughly evaluate the expected
contribution of Compton scattering by the non-thermal electrons, which
could produce X-ray and $\gamma$-ray photons from seed photons of the
synchrotron radiation (synchrotron self Compton; SSC) and external
photons (external Compton; EC) from the accretion disk, respectively, 
for a typical Lorentz factor of $\gamma = 40$.
The energy density of synchrotron photons in the near infrared to optical bands
is estimated as
$$
u_{\rm sync} \sim L_{\rm sync}/(4 \pi R^2 c) \approx 7\times10^6 {\rm \;\;erg\; cm}^{-3},
$$
where $L_{\rm sync}$, which depends on 
$\gamma_{\rm max}$, is assumed to be $10^{36}$ erg s$^{-1}$.
Similarly, that of external photons observed
in the X-ray to Gamma-ray bands is 
$$
u_{\rm Comp} \sim L_{\rm Comp}/(4 \pi l^2 c) \approx 2\times10^6 {\rm \;\;erg\; cm}^{-3},
$$ where $L_{\rm Comp} \sim 2\times10^{37}$ erg s$^{-1}$ and the
distance from the disk and Comptonized corona to the jet base is assumed to be $l =
5\times10^{9}$ cm that corresponds to the time delay ($\approx 150$ ms)
between the optical and X-ray cross correlation signals
\citep{gan08}. 

Thus, the photon energy density of these seed
photons is $\approx$10--40 times smaller than that of the magnetic field, $u_B =
8.4\times10^7$ erg cm$^{-3}$.  As noticed from Figure~\ref{SED}, recalling
that the ratio of the luminosity between synchrotron and SSC 
by the same electrons is proportional to $u_B/u_{\rm sync}$, 
we estimate that the contribution of Comptonized
photons observed in the 1--100 keV X-ray band (SSC) should
be approximately 0.4\% of the total observed luminosity, which is dominated
by thermal Comptonization by the hot corona. This justifies our X-ray
spectral model, which ignores any non-thermal emission from the jets.
The EC component, however, may contribute to the high energy emission
above $\sim$1 MeV.

Similarly, we also ignore the jet synchrotron emission as seed
photons for thermal Comptonization by the corona in our model, because
the emitting regions in the jets are assumed to be distant from the
corona and disk. However, we cannot rule out models where the corona and
jet base are co-spatial and a Comptonized component of the synchrotron
photons significantly contributes to the X-ray emission, as discussed in
\citet{mar05}. We leave it for future work to apply self-consistent jet
and corona models to the simultaneous, multi-wavelengths SED including
Suzaku data.

\begin{figure}
\begin{center}
\FigureFile(80mm,){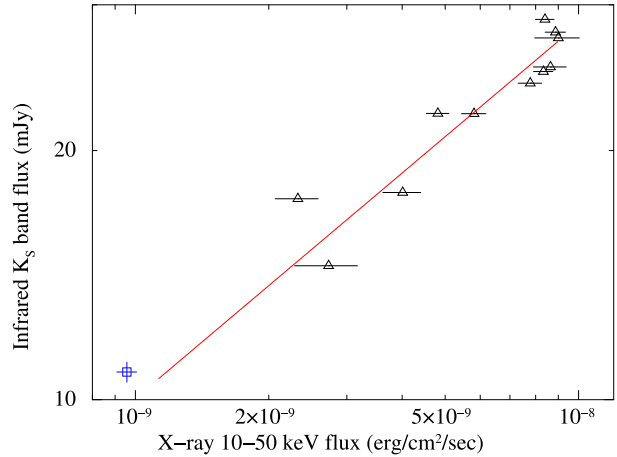}
\caption{The relation between the X-ray (15--50 keV) and IR ($K_{\rm S}$)
fluxes of GX 339--4 with Swift/BAT and IRSF/SIRIUS in 2009
March. The result of the 1997 observation \citep{cor02} is also
included (blue, open square). Each $K_{\rm S}$ band flux is calculated by averaging all
the magnitudes obtained on the same day. The data points
except for the 1997 data are fitted by a power law with 
$F_{\rm Ks} \propto F_{\rm X}^\mathit{\Gamma}$). The best-fit result
($\mathit{\Gamma}=0.45$) is shown as a solid line in this figure.
\label{fig_flux_add}}
\end{center}
\end{figure}

To examine the correlation between the $K_{\rm s}$ and hard X-ray
light curves (Figure \ref{lc_irX}), we plot the one-day averaged 15--50
keV flux ($F_{\rm X}$) versus $K_{\rm s}$ band flux ($F_{\rm Ks}$) in
Figure~\ref{fig_flux_add}. The data point taken from the 1981 dataset
\citep{cor02} is also plotted. By fitting IRSF and
Swift points with a power law of $F_{\rm Ks} \propto F_{\rm
X}^{\mathit{\Gamma}}$, we obtain $\mathit{\Gamma} = 0.45\pm0.06$. This value is close to the
\citet{cor09} result from GX 339--4 ($0.47 \leq \mathit{\Gamma} \leq 0.49$),
which was obtained between the $H$ and 3--9 keV bands. 
Similar correlation is reported between the radio and X-ray fluxes with a
steeper slope of 0.7--0.9 in the low/hard state\citep{cor03}. 
These correlations confirm the strong link between
the jet formation and accretion flow, as directly indicated by the
strong cross correlation between the optical and X-ray bands on much
shorter ($<1$ sec) time scale for this source. 

The slope of the correlation and its dependence on the wavelength and
luminosity give us important clues to understand the origin of the
SED. As mentioned above, our result is not compatible with models
where the near infrared emission is assumed to be an optically thick
synchrotron emission. We note that, however, the thermal emission
from the irradiated disk must be taken into account as well. Its
spectra is represented as a superposition of blackbody radiation that
results in a flat SED if the conversion efficiency from the irradiated
flux is constant over the whole disk \citep{fra02}. The relation between
the incident luminosity ($L_{\rm X}$) and that of thermal emission
from an irradiated outer disk ($L_{\nu,{\rm irr}}$) is given by
$L_{\nu,{\rm irr}}\propto L_{\rm X}$ at frequencies corresponding to
the flat part, and $L_{\nu,{\rm RJ}}\propto L_{\rm X}^{1/4}$ in the
Rayleigh-Jeans region (see \cite{cor09}). Thus, it is possible that
a contribution from the irradiate disk in the Rayleigh-Jeans part
could make the observed slope $\Gamma$ somewhat flatter than
those expected from jet models. To prove the origins of the near IR
emission, detailed modeling of the complete SED would be necessary,
which we leave for our future investigation.

\section{Conclusion}

We observed GX 339--4 with Suzaku and IRSF in the low/hard
state at $\sim 2$\% of the Eddingtion luminosity. The conclusions are
summarized as follows.

\begin{enumerate}

\item 
The broad-band Suzaku spectra are well represented by a thermal
Comptonization by hot $\sim 200$ keV electrons of seed photons from
the optically thick disk, with a small contribution of the direct disk
component. The corona has at least two different optical depths,
$\tau\approx$ 0.4 and 1, indicating inhomogeneous structure.

\item Analysis of the iron-K line with a diskline model yields an
innermost radius of $R_{\rm in}=(13.3^{+6.4}_{-6.0}) R_{\rm g}$ with an
estimated inclination of $\approx 50^\circ$.

\item The Suzaku results indicate that the optically thick
accretion disk is truncated before the ISCO and its inner regions are
almost fully covered by the hot corona. The inferred innermost radii
from the continuum fit and the iron-K line profile are consistent each
other.
 
\item The near infrared fluxes are correlated with that of hard X-rays
by the relation $F_{\rm Ks} \propto F_{\rm X}^{0.45}$, with variable
spectral indices from --0.20 to +0.08. The spectrum can be explained
by a sum of an optically thin synchrotron emission and a thermal
emission from the X-ray irradiated outer disk. We estimate the
magnetic field and size of the jet base to be $5\times10^4$ G and
$6\times10^8$ cm, respectively.

\end{enumerate}

\bigskip

This work was partly supported by the
Grant-in-Aid for Scientific Research 20540230 (YU), and by the
grant-in-aid for the Global COE Program ``The Next Generation of
Physics, Spun from Universality and Emergence'' from the Ministry of
Education, Culture, Sports, Science and Technology (MEXT) of Japan.
We are grateful to Suzaku operation team for carrying out the ToO
observations.

\end{document}